\documentclass{aa}
\usepackage{graphicx}
\usepackage{txfonts}
\usepackage{natbib}  
\usepackage{subcaption}
\usepackage{color}
\usepackage{ulem}
\usepackage{amsmath}  

\begin{document} 
 
   \title{Sodium abundances of AGB and RGB stars in galactic globular clusters}

   \subtitle{I. Analysis and results of NGC\,2808 
             \thanks{Based on observations made with ESO telescopes at the La Silla Paranal Observatory under programme ID 093.D-0818(A).}}

   \author{Y. Wang 
           \inst{1,2,3}
           \and
           F. Primas \inst{1}
           \and
           C.Charbonnel \inst{4,5} 
           \and
           M. Van der Swaelmen \inst{6} 
           \and
           G. Bono \inst{7,8}
           \and
           W. Chantereau \inst{4}
           \and
           G. Zhao \inst{2}
          }

   \institute{European Southern Observatory (ESO), Karl-Schwarschild-Str. 2, 85748, Garching b. München, Germany \\
              \email{fprimas@eso.org}
         \and
             Key Laboratory of Optical Astronomy, National Astronomical Observatories, Chinese Academy of Science, Beijing 100012, China \\
              \email{ywang@nao.cas.cn; gzhao@nao.cas.cn}
         \and
             School of Astronomy and Space Science, University of Chinese Academy of Sciences, Beijing 100049, China \\
         \and
             Department of Astronomy, University of Geneva, Chemin des Maillettes 51, 1290, Versoix, Switzerland \\
              \email{corinne.charbonnel@unige.ch}
         \and
             IRAP, UMR 5277 CNRS and Université de Toulouse, 14 Av. E. Belin, F-31400 Toulouse, France \\
         \and
             Institut d’Astronomie et d’Astrophysique, Université Libre de Bruxelles, CP. 226, Boulevard du Triomphe, 1050 Brussels, Belgium \\
         \and
             Dipartimento di Fisica, Università di Roma Tor Vergata, via della Ricerca Scientifica 1, 00133 Rome, Italy \\
         \and
             INAF – Osservatorio Astronomico di Roma, via Frascati 33, Monte Porzio Catone, Rome, Italy \\
             }
             
   \date{accepted May 23, 2016}

 
  \abstract
   {Galactic globular clusters (GC) are known to have multiple stellar populations and be characterised by similar chemical
   features, e.g. \element{O}$-$\element{Na} anti-correlation. While second-population stars, identified by their \element{Na} 
   overabundance, have been found from the main sequence turn-off up to the tip of the red giant branch in various Galactic GCs, 
   asymptotic giant branch (AGB) stars have rarely been targeted. The recent finding that NGC\,6752 lacks an \element{Na}-rich 
   AGB star has thus triggered new studies on AGB stars in GCs, since this result questions our basic understanding of GC 
   formation and stellar evolution theory. 
   }
   {We aim to compare the \element{Na} abundance distributions of AGB and RGB stars in Galactic GCs and investigate whether 
   the presence of \element{Na}-rich stars on the AGB is metallicity-dependent. 
   }
   {With high-resolution spectra obtained with the multi-object high-resolution spectrograph FLAMES on ESO/VLT, we derived 
   accurate \element{Na} abundances for 31 AGB and 40 RGB stars in the Galactic GC NGC\,2808. 
   }
   {We find that NGC\,2808 has a mean metallicity of $-1.11\pm0.08\,\mathrm{dex}$, in good agreement with earlier analyses. 
   Comparable \element{Na} abundance dispersions are derived for our AGB and RGB samples, with the AGB stars being slightly 
   more concentrated than the RGB stars. The ratios of \element{Na}-poor first-population to \element{Na}-rich second-population 
   stars are 45:55 in the AGB sample and 48:52 in the RGB sample. 
   }
   {NGC\,2808 has \element{Na}-rich second-population AGB stars, which turn out to be even more numerous $-$ in relative terms $-$ 
   than their \element{Na}-poor AGB counterparts and the \element{Na}-rich stars on the RGB. Our findings are well reproduced by 
   the fast rotating massive stars scenario and they do not contradict the recent results that there is not an \element{Na}-rich 
   AGB star in NGC\,6752. NGC\,2808 thus joins the larger group of Galactic GCs for which \element{Na}-rich second-population stars 
   on the AGB have recently been found.}


  \maketitle


\section{Introduction}
\label{section:introduction}

Galactic globular clusters (GC) have been subjected to extensive studies of their chemical characteristics for several decades. 
In the early 90s, the dedicated Lick-Texas spectroscopic survey of several GCs paved the way and discovered that oxygen and sodium abundances 
in red giant branch (RGB) stars anti-correlate (e.g. the series by \citealp{Kraft1992, Sneden1992, Kraft1993, Sneden1994, Kraft1995}). Later on, 
the advent of more efficient single/multi-object spectrographs, mounted on $8-10\,\mathrm{m}$ class telescopes, allowed for more systematic studies 
of larger stellar samples down to the turn-off and along the main sequence. Twenty-five years later, the \element{O}$-$\element{Na} anti-correlation 
is recognized as being a chemical feature common to most (if not all, though at different degrees of significance) Galactic GCs \citep{Carretta2010}. 

This feature is interpreted as the proof of the existence of (at least) two stellar populations (often referred to as different stellar generations) 
co-inhabiting the cluster. While first-population (1P) GC stars display \element{Na} and \element{O} abundances consistent with that of halo field 
stars of similar metallicity, second-population (2P) stars can be identified by their \element{Na} overabundances and \element{O} deficiencies that 
are associated with other chemical peculiarities (e.g. nitrogen and aluminium enrichment, and carbon, lithium, and magnesium deficiency; 
e.g. \citealp{Pancino2010,Villanova2011,Carretta2014a,Carretta2014b,Lapenna2015}). First- and second-population stars are also related to the 
appearance of multimodal sequences in different regions of GC colour-magnitude diagramme (CMD; e.g. \citealp{Piotto2012, Piotto2015, Milone2015a, 
Milone2015b, Nardiello2015}), which can (at least partly) be associated with helium abundance variations in their initial chemical composition 
(see e.g. \citealp{Chantereau2015} and references therein). 
All these pieces of evidence point to all GCs having suffered from self-enrichment during their early evolution, where 2P stars formed out of the 
\element{Na}-rich, \element{O}-poor ashes of hydrogen burning at high temperature ejected by more massive 1P stars and diluted with interstellar gas 
(e.g. \citealp{PrantzosCharbonnel2006, Prantzos2007}). However, the nature of the polluters remains highly debatable, as well as the mode and 
timeline of the formation of 2P stars. As of today, none of the proposed models for GC early evolution is able to account for the  chemical features 
that are common to all GCs, nor for the spectrocopic and photometric diversity of these systems (e.g. \citealp{Bastian2015,Renzini2015,Krause2016}).
This raises serious challenges related to our understanding of the formation and evolution of massive star clusters and of galaxies in a more general 
cosmological context.
 
Therefore, because of the importance of these specific chemical features (like the \element{O}$-$\element{Na} anti-correlation), a wealth of 
observational data has been gathered for a sizable number of Galactic GCs. 
Cluster stars have been observed and analysed at different evolutionary phases, down to the main sequence, although the majority 
of the data is from the brighter areas of the CMDs (RGB and HB stars, in particular). Thanks to these analyses, it has 
been possible to prove the existence of the \element{O}$-$\element{Na} anti-correlation at all evolutionary phases in GCs: 
from the main sequence (MS; e.g. \citealp{Gratton2001, Lind2009, DOrazi2010, Monaco2012, Dobrovolskas2014}) and the sub giant branch (SGB; 
e.g. \citealp{Carretta2005, Lind2009, Pancino2011, Monaco2012}) to the red giant branch (RGB; e.g. \citealp{Yong2013, Cordero2014, Carretta2014a, 
Carretta2015} and \citealp{Gratton2012AARv} and references therein) and the horizontal branch (HB; e.g. \citealp{Villanova2009} and the series 
by \citealp{Gratton2011, Gratton2012, Gratton2013, Gratton2014, Gratton2015}). 

However, despite this large number of investigations, asymptotic giant branch (AGB) stars have rarely been targeted systemically because of their paucity 
in GCs (a result of their short lifetime). \citet{Pilachowski1996} studied a few of them in M\,13 (NGC\,6205; 112 RGBs, and 18 AGBs) and found them to be 
rich in sodium ([\element{Na}/\element{Fe}] $>0.05\,\mathrm{dex}$), similar to the RGB-tip stars ($\log g<1$) but with a slightly lower overall \element{Na} 
abundance. \citet{Johnson2012} looked again at the same cluster (M\,13) and derived \element{Na} and \element{O} abundances for 98 RGB and 
$\sim$\,15 AGB stars, finding very similar results to \citet{Pilachowski1996}, since 66 RGB and twelve AGB stars are in common with the 1996 sample. 
Since no extreme (very \element{O}-poor) AGB star is present in their sample, the authors concluded that only the most \element{Na}-rich 
and \element{O}-poor stars may have failed to reach the AGB. 

More recently, a spectroscopic study by \citet{Campbell2013} revealed the lack of \element{Na}-rich, 2P stars along the early-AGB of NGC\,6752. 
This came as a surprise since, in this GC, as well as all the Milky Way GCs studied so far, 1P and 2P stars have been found at 
the MS turn-off, on the SGB and on the RGB, and the \element{Na}-rich, 2P stars are even twice as numerous as their 1P counterparts 
\citep{PrantzosCharbonnel2006,Carretta2010}. It was therefore concluded by Campbell and collaborators that in NGC\,6752 only 1P stars manage 
to climb the AGB, possibly posing a fundamental problem to stellar evolution theory. This result then triggered a new study of 35 AGB stars 
in the Galactic GC 47\,Tuc (NGC\,104, \citealp{Johnson2015}), which found that, in contrast to NGC\,6752, the AGB and RGB populations of 
47\,Tuc have nearly identical [\element{Na}/\element{Fe}] dispersions, with only a small fraction ($\lesssim20\%$) of \element{Na}-rich stars 
that may fail to ascend the AGB, which is similar to what was observed in M\,13. A new study of 6 AGB and 13 RGB stars in M\,62 
(NGC\,6266; \citealp{Lapenna2015}) find their AGB stars to behave similarly to what was found in NGC\,6752, i.e. they are all \element{Na}-poor 
and \element{O}-rich (1P) stars. On the other hand, \citet{GarciaHernandez2015} clearly show that 2P AGB stars exist in metal-poor GCs, with a 
study of \element{Al} and \element{Mg} abundances in 44 AGB stars in four metal-poor GCs (M\,13, M\,5, M\,3, and M\,2). 
Therefore, the question of the presence of 2P AGB stars in GCs with various properties (e.g. different ages, metallicities, etc) is far from being 
settled, although it might bring interesting constraints on the self-enrichment mechanisms (e.g. \citealt{CCWC2016}).

Considering how limited the current sample of GC AGB stars is in terms of \element{Na} (and \element{O}) abundance determinations, we embarked 
on a new observational campaign to increase the number of AGB stars for which accurate \element{Na} abundances can be derived. We started by 
observing RGB and AGB stars in NGC\,2808, a moderately metal-poor Galactic GC, well known for its multiple stellar populations (e.g. chemically: 
\citealp{Carretta2006}; photometrically: \citealp{Piotto2007}). Thanks to the large number of data available in the literature for NGC\,2808, 
our first goal is to use NGC\,2808 as our test-bench cluster, to establish our analysis procedures, before applying the same methodology 
to a larger number of clusters, spanning a range of metallicities. 

The paper is organised as follows: Sections 2 and 3 describe in detail the observations and the analysis of the data; Section 4 presents our derived 
\element{Na} abundance of NGC\,2808; the discussion and summary in Sects. 5 and 6 close the paper and suggest future steps; finally, in the 
Appendix we discuss  the influence of different methods to determine the stellar parameters on the derived \element{Na} abundances. 

\section{Observations and data reduction}

Our targets were selected from the Johnson-Morgan photometric database that is part of the project described in \citet{Stetson2000, Stetson2005}, 
and cover a magnitude range of about 1.5 magnitudes ($\mathrm{V}=15.1-13.5$\,mag). A total of 53 AGB stars and 47 RGB stars were selected and 
observed with the high-resolution multi-object spectrograph FLAMES, mounted on ESO/VLT-UT2 \citep{Pasquini2003}. For our programme, we used 
FLAMES in combined mode, i.e. we observed simultaneously the brightest five objects of our sample with UVES-fibre and the remaining targets 
with GIRAFFE/Medusa. For UVES-fibre, we chose the Red 580 setting, whereas for GIRAFFE we selected the HR\,13, HR\,15, and HR\,19 set-ups. 
More details about the observations, which were carried out in service mode, are summarised in Table \ref{obslog}. 

\begin{table} \small
\caption{Log of the observations for NGC\,2808}              
\label{obslog}
\centering
\begin{tabular}{c c c c c c }
\hline\hline                 
 Instrument   & Set-up   &  $R$  & $\lambda$-range & Exp.time         \\
              &          &       &   (nm)          & (s)              \\
\hline
 GIRAFFE      & HR\,13   & 22500 & 612$-$640.5     & 4$\times$3600        \\
              & HR\,15   & 19300 & 660.7$-$696.5   & 4$\times$2700        \\
              & HR\,19   & 14000 & 774.5$-$833.5   & 4$\times$3600        \\
 UVES$-$fibre & Red\,580 & 47000 & 480$-$680       & 8$\times$3600 and    \\
              &          &       &                 & 4$\times$2700        \\
\hline
\end{tabular}
\end{table}

Primary data reduction (including the bias correction, wavelength calibration using a Th-Ar lamp, spectrum extraction, and flat fielding) was performed 
with the ESO GIRAFFE and UVES pipelines, respectively. Sky-subtraction (by averaging seven sky fibres for GIRAFFE sample and one for UVES sample), 
radial velocity measurement and correction were applied at the end of the reduction procedure. For the GIRAFFE HR\,13 and HR\,19 spectra, we also 
performed the telluric correction using the recently released ESO Sky Tool Molecfit (see \citealp{Smette2015} and \citealp{Kausch2015} for more details). 
Finally the spectra from the same setup and for the same stars were co-added achieving signal-to-noise ratios (S/N) that range from 100 to 350 for 
GIRAFFE spectra and 90$-$180 for UVES spectra, depending on the magnitude of the star. 

\begin{figure}
    \centering
    \includegraphics[width=0.42\textwidth]{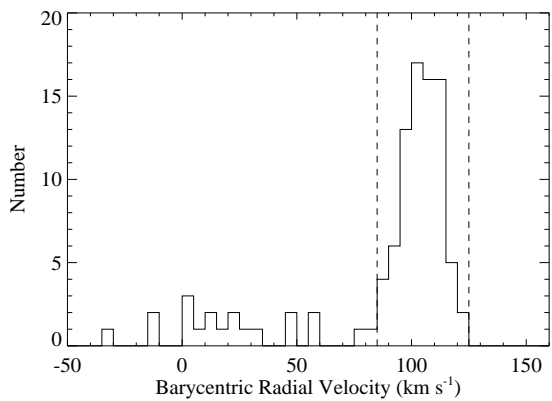}
    \caption{Barycentric radial velocity distribution. 
             The two vertical dashed lines mark the radial velocity range where we select stars as cluster members.}
    \label{RV}
  \end{figure}

The barycentric corrections for the radial velocities were derived with the ESO hourly airmass tool\footnote{Web: http://www.eso.org/sci/observing/tools/calendar/airmass.html}. 
Figure \ref{RV} shows the barycentric radial velocity distribution, as derived for all our observed stars. Although the main peak of the distribution is 
populated by the majority of our initial sample stars, a long tail towards smaller velocities is also present. We thus only considered as cluster members 
those stars belonging to the main peak (i.e. all those falling between the two vertical dashed lines in Fig. \ref{RV}, with barycentric radial velocities 
of roughly between 85 and 125$\,\mathrm{km\,s^{-1}}$). From this group, we had to further exclude six objects (because of too high metallicity, \ion{Fe}{i} 
and \ion{Fe}{ii} abundance mismatches that were incompatible with a reasonable derivation of the reddening, and spectroscopic binary). Our final sample thus 
consists of 73 stars in total, 33 AGB stars, and 40 RGB stars. Of these, five objects (three AGB and two RGB stars) were observed with UVES$-$fibre (the UVES 
sample hereafter) and 68 objects (30 AGB and 38 RGB stars) were observed with FLAMES/GIRAFFE (the GIRAFFE sample hereafter). The mean barycentric radial 
velocity obtained from all the cluster members is $104.6\,\mathrm{km}\,\mathrm{s}^{-1}$ with a dispersion of $\sigma=8.0\,\mathrm{km}\,\mathrm{s}^{-1}$, 
in good agreement with \citet{Carretta2006} ($102.4\,\mathrm{km}\,\mathrm{s}^{-1}$ with $\sigma=9.8\,\mathrm{km}\,\mathrm{s}^{-1}$). 

Table \ref{basic} lists the evolutionary phase (AGB/RGB), instrument used for collecting the spectrum (UVES/GIRAFFE), coordinates, photometry, and 
barycentric radial velocities of our member stars. Their location in the CMD is shown in Fig.\ref{CMD}.

\begin{sidewaystable*} \scriptsize
\caption{Basic information of our sample stars: evolutionary phase, instrument used for observation, coordinates, photometry, and barycentric radial 
         velocity. The complete table is available electronically; we show here the first line for guidance.}       
\label{basic}
\centering
\begin{tabular}{c c c c c c c c c c c c c c c c c c c c}
\hline\hline                 
 Star ID$^{a}$ & Evol. Ph. & Instrument & RA (J2000) & Dec (J2000) &  B    &   e\_B &   V    &   e\_V  &   I    &   e\_I  & J\_2mass &  e\_J & H\_2mass &  e\_H & K\_2mass &  e\_K & RV($\mathrm{km\,s}^{-1}$)  \\  
\hline
 AGB46601  & AGB &  UVES   &  09 11 30.77  &  -64 53 37.90  &  15.369  &  0.0010  &  13.998  &  0.0009  &  12.520  &  0.0014  &  11.430  &  0.023  &  10.676  &  0.022  &  10.540  &  0.023  &  109.990   \\ 
$\dots$ & $\dots$ & $\dots$ & $\dots$ & $\dots$ & $\dots$ & $\dots$ & $\dots$ & $\dots$ & $\dots$ & $\dots$ & $\dots$ & $\dots$ & $\dots$ & $\dots$ & $\dots$ & $\dots$ & $\dots$  \\  
\hline     
\end{tabular}
\tablefoot{$^{a}$ Star ID reports the original ID from the photometric catalogue, to which we added the suffix AGB/RGB to ease our own data handling.}
\end{sidewaystable*}

\section{Stellar parameters and abundance analysis} 

\subsection{Effective temperature and surface gravity}

We used the photometric method to derive the stellar effective temperature ($T_\mathrm{eff}$) and surface gravity ($\log g$). 
Optical B, V, and I magnitudes are available for all our stars. By cross-matching the coordinates of our targets with the 2MASS catalogue 
\citep{Skrutskie2006}, we have also been able to extract the infrared J, H, and K magnitudes for all except six stars (for which no 2MASS 
counterpart could be identified). 

We adopted the reddening value of $\mathrm{E(B-V)}=0.22\,\mathrm{mag}$ (\citealp{Harris1996}, 2010 edition), together with 
the \citet{Cardelli1989} relations: 
\begin{equation*}
\begin{split}
  \mathrm{A(B)} &= 4.145\,\mathrm{E(B-V)} \\
  \mathrm{A(V)} &= 3.1\,\mathrm{E(B-V)}   \\
  \mathrm{A(I)} &= 1.485\,\mathrm{E(B-V)} \\
  \mathrm{A(J)} &= 0.874\,\mathrm{E(B-V)} \\
  \mathrm{A(H)} &= 0.589\,\mathrm{E(B-V)} \\
  \mathrm{A(K)} &= 0.353\,\mathrm{E(B-V)}.
\end{split}
\end{equation*}

We used the \citet{RamirezMelendez2005} photometric calibrations for giants (which are provided as a function of the colour index and 
[\element{Fe}/\element{H}]) and computed five scales of photometric temperatures, using the de-reddened $\mathrm{(B-V)}_{0}$, $\mathrm{(V-I)}_{0}$, 
$\mathrm{(V-J)}_{0}$, $\mathrm{(V-H)}_{0}$, and $\mathrm{(V-K)}_{0}$ colour indices (see Table \ref{dTeff} for a comparison between the five scales). 
The mean value of the five temperature scales ($T_\mathrm{eff,mean}$) was adopted as our final effective temperature except for the six stars without 
J, H, and K magnitudes for which we use the mean relation $T_\mathrm{eff,mean}$ {\it vs.} $T_\mathrm{eff,(V-I)}$, as derived for the rest of the sample. 

\begin{table}
\caption{Mean differences between photometric temperature scales and standard deviations.}      
\label{dTeff}      
\centering                          
\begin{tabular}{c c c c c}        
\hline\hline                 
       & $T_\mathrm{eff,(V-I)}$ & $T_\mathrm{eff,(V-J)}$ & $T_\mathrm{eff,(V-H)}$ & $T_\mathrm{eff,(V-K)}$  \\    
\hline                        
  $T_\mathrm{eff,(B-V)}$ & $-3\pm38$ & $131\pm46$ & $162\pm43$ & $163\pm51$   \\      
  $T_\mathrm{eff,(V-I)}$ &     -     & $134\pm41$ & $165\pm36$ & $166\pm41$   \\
  $T_\mathrm{eff,(V-J)}$ &     -     &     -      &  $31\pm30$ &  $32\pm40$   \\
  $T_\mathrm{eff,(V-H)}$ &     -     &     -      &     -      &   $1\pm31$   \\
\hline                                   
\end{tabular}
\end{table}

To evaluate the error on our final set of effective temperatures, we took into account four main sources of uncertainty: 
the dispersion $\sigma_\mathrm{cal}$ of the photometric calibration itself (taken from \citealp{RamirezMelendez2005}, Table 3; smaller than 
$40\,\mathrm{K}$ for the colours we used); the differential reddening ($0.02\,\mathrm{mag}$ in $\mathrm{E(B-V)}$, \citealp{Bedin2000}); 
the uncertainty in the colour index $\sigma_\mathrm{colour}$ and on the [\element{Fe}/\element{H}] ratio. 
After propagating all errors, we ended up with a typical error on the final $T_\mathrm{eff}$ of about $\pm70\,\mathrm{K}$. The largest contributing 
source is the differential reddening ($\sim55\%$), followed by the $\sigma_\mathrm{cal}$ of the calibration itself ($\sim33\%$) and the error on 
the magnitudes ($\sim11\%$), while the error on the derived metallicity is negligible (about $1\%$).  

\begin{figure}
    \centering
    \includegraphics[width=0.47\textwidth]{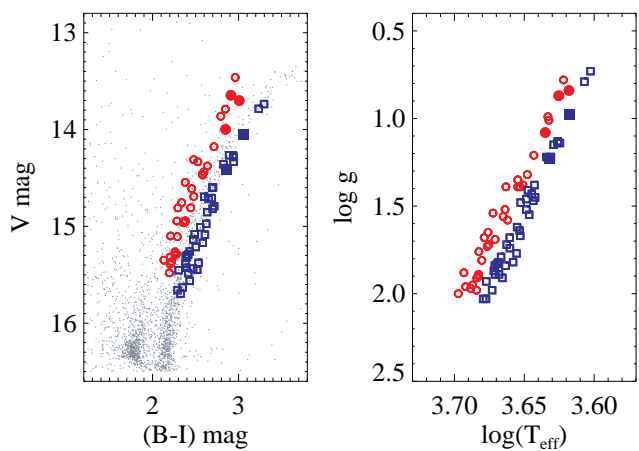}
    \caption{Photometric CMD and $\log g - \log(T_{\rm eff})$ distribution. 
             The left panel shows the CMD of the cluster member star sample; 
             and the right panel shows their $\log g - \log(T_{\rm eff})$ distribution.
             The red circles and blue squares represent AGB and RGB stars, respectively, 
             while the GIRAFFE and UVES samples are distinguished by open and filled symbols, respectively.
             These symbols are used through out this paper.}
    \label{CMD}
  \end{figure}

The surface gravities $\log g$ were derived from effective temperatures and bolometric corrections, assuming that the stars have masses of $0.85\,M_{\odot}$, 
as adopted by \citet{Carretta2006}. The bolometric corrections of our stars were obtained following the relations by \citet{Alonso1999}. We adopted a 
visual distance modulus of $15.59\,\mathrm{mag}$ (\citealp{Harris1996}, 2010 edition) and the bolometric magnitude of the Sun of $M_\mathrm{bol,\odot}=4.75$. 
The typical error on $\log g$ is about $\pm0.05\,\mathrm{dex}$, strongly dominated by the uncertainties in effective temperature ($\sim61\%$) and the 
differential reddening ($\sim39\%$). The right panel of Fig. \ref{CMD} shows the final $\log g - \log(T_\mathrm{eff})$ distribution of the member stars. 

\subsection{Metallicity and microturbulent velocity}
\label{section:Metallicity}

To derive the metallicity [\element{Fe}/\element{H}] of our stars, we selected unblended \ion{Fe}{i} and \ion{Fe}{ii} lines from the VALD3 
database\footnote{Web interface at http://vald.inasan.ru/~vald3/php/vald.php} \citep{Piskunov1995VLAD, Kupka2000VALD, Ryabchikova2015VALD} where 
the atomic data was originally taken from \citet{K07, FMW, BWL, BKK, BK} for \ion{Fe}{i} lines and from \citet{K13} and \citet{BSScor} for \ion{Fe}{ii} 
lines, and optimised our selection for GIRAFFE and UVES spectra separately according to their different spectral resolutions and wavelength coverages. 
The equivalent widths (EWs) of the spectral lines were measured using the automated tool DAOSPEC \citep{Stetson2008}, which also outputs the error 
associated to the derivation of each line equivalent width. To keep the determination of the metallicity as accurate as possible, we excluded very weak 
($\leq20\,\mathrm{m}$\AA) and very strong ($\geq120\,\mathrm{m}$\AA) iron lines. We used 1D LTE spherical MARCS model atmospheres \citep{Gustafsson2008} 
and the LTE stellar line analysis programme MOOG (\citealp{Sneden1973}, 2014 release) to derive the metallicity and the microturbulence velocity 
($\xi_\mathrm{t}$), the latter by requiring \ion{Fe}{i} abundances to show no trend with the reduced equivalent widths ($\log(W_{\lambda}/\lambda)$) of 
the lines. An iterative procedure was then followed to consistently derive all stellar parameters ($T_\mathrm{eff}$, $\log g$, [\element{Fe}/\element{H}],
and $\xi_\mathrm{t}$), because of their known interdependences. 

We ended up using typically $20-35$ \ion{Fe}{i} and $4-5$ \ion{Fe}{ii} lines for the GIRAFFE spectra and $50-65$ \ion{Fe}{i} and $6-8$ \ion{Fe}{ii} 
lines for the UVES spectra. The solar iron abundance of $\log\epsilon\mathrm{(Fe)}_{\odot}=7.50$ from \citet{Asplund2009} is adopted throughout our 
analysis. Tables \ref{stellarpar} and \ref{metal} summarise our final stellar parameters and average iron abundances, respectively.

Our derived iron abundances are consistent with those found in the literature. (see Table \ref{metal}). We find a small difference between 
the [\ion{Fe}{i}/\element{H}] and [\ion{Fe}{ii}/\element{H}] ratios, which disappears once we correct the \ion{Fe}{i} values for non-LTE corrections 
(see Sect. \ref{section:NLTE}). Therefore, we derive an overall metallicity for NGC\,2808 of [\ion{Fe}/\element{H}]$=-1.11\,\mathrm{dex}$ 
($\mathrm{rms}=0.09\,\mathrm{dex}$). 

\begin{table*}  \small
\caption{Stellar parameters of our sample stars. 
         The complete table is available electronically; we show here the first line for guidance.}      
\label{stellarpar}      
\centering              
\begin{tabular}{c c c c c c c c c c c}    
\hline\hline        
 Star ID  & Evol. Ph. & Instrument & $T_\mathrm{eff}$ & $\log g$ & $\xi_{\rm t}$ & [\ion{Fe}{i}/\element{H}]$_\mathrm{LTE}$ & rms\_lines & [\ion{Fe}{ii}/\element{H}] & rms\_lines & [\ion{Fe}{i}/\element{H}]$_\mathrm{NLTE}$   \\ 
          &      &            & ($\mathrm{K}$)  &      & ($\mathrm{km\,s}^{-1}$) & ($\mathrm{dex}$) & ($\mathrm{dex}$) & ($\mathrm{dex}$) & ($\mathrm{dex}$) & ($\mathrm{dex}$)    \\ 
\hline     
  AGB46601  & AGB &  UVES   &  4315  &   1.08  &   1.73  &  -1.14  &   0.09  &  -1.06  &   0.04  &  -1.10   \\
$\dots$ & $\dots$ & $\dots$ & $\dots$ & $\dots$ & $\dots$ & $\dots$ & $\dots$ & $\dots$ & $\dots$ & $\dots$  \\ 
\hline       
\end{tabular}
\end{table*}

\begin{table*}  \small
\caption{Metallicity of NGC\,2808 from this work and literature.}       
\label{metal}      
\centering         
\begin{tabular}{c c c c c c c c}     
\hline\hline        
 [\ion{Fe}{i}/\element{H}]$_\mathrm{LTE}$ &  rms  & n$_\mathrm{star}$ & [\ion{Fe}{ii}/\element{H}] &  rms  & n$_\mathrm{star}$ & Evol. Ph. & Reference    \\
      ($\mathrm{dex}$)             & ($\mathrm{dex}$) &               &    ($\mathrm{dex}$)   & ($\mathrm{dex}$) &             &             &            \\
\hline     
    -1.19   &  0.10   &  33   &  -1.14   &  0.10   &  33   &  AGB   &  This work                \\
    -1.12   &  0.07   &  40   &  -1.09   &  0.07   &  40   &  RGB   &  This work                \\
    -1.14   &  0.06   &  19   &  -1.14   &  0.13   &  19   &  RGB   &  \citet{Carretta2004}     \\
    -1.10   &  0.07   & 123   &  -1.16   &  0.09   &  90   &  RGB   &  \citet{Carretta2006}     \\
    -1.15   &  0.08   &  12   &  -1.18   &  0.09   &  12   &  RGB   &  \citet{Carretta2009b}    \\
\hline              
\end{tabular}
\tablefoot{The 12 stars in \citet{Carretta2009b} are selected from the sample of \citet{Carretta2006}, and there are two stars in common  
           between the samples of \citet{Carretta2004} and \citet{Carretta2006}.}
\end{table*}

\subsection{Sodium abundances}
\label{section:Na}

Despite the presence of three different \ion{Na}{i} doublets in our spectra ($6154-6160$\,\AA\ in both UVES and GIRAFFE spectra, 
$8183-8194$\,\AA\ only in the GIRAFFE spectra, and $5682-5688$\,\AA\ only in the UVES spectra), we were able to reliably use only the doublet in 
common to all spectra (i.e. $6154-6160$\,\AA) because the other two show saturation at the metallicity of NGC\,2808. This doublet also has a small 
drawback, namely the $6160$\,\AA\ line blends with a calcium line, but this can be overcome by analysing the Na doublet via spectrum synthesis. 

\begin{figure}
    \centering
    \includegraphics[width=0.4\textwidth]{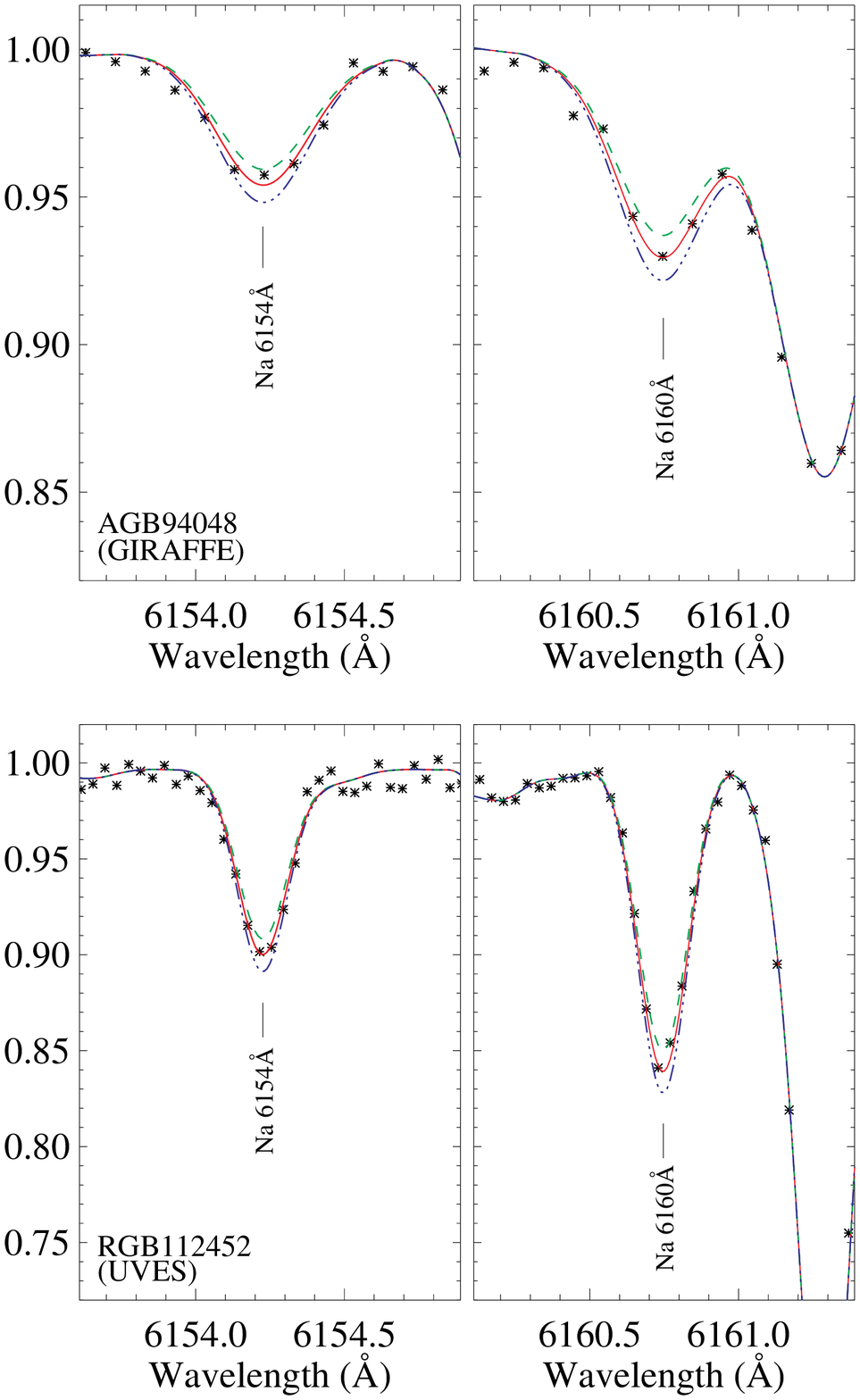}
    \caption{Spectra syntheses of the \element{Na} lines region. 
             The top panel shows the spectra of the GIRAFFE AGB star AGB94048, 
             and the panel shows the spectra of the UVES RGB star RGB star RGB112452.
             The black asterisks represent the observed spectra; the red solid lines are the best-fit synthesised spectra; 
             the blue dash-dotted lines and the green dashed lines are synthesised spectra, but with the best-fit \element{Na} 
             abundance changed by plus/minus the error of \element{Na} abundance ($\pm0.07\,\mathrm{dex}$ for AGB94048 and 
             $\pm0.05\,\mathrm{dex}$ for RGB112452; here the error is a combination of random error and fitting error). 
             }
    \label{Spec_Na}
  \end{figure}

For this purpose, we used MOOG and our interpolated suite of MARCS model atmospheres, matching our derived stellar parameters. 
The atomic data of the \element{Na} doublet was adopted from the VALD3 database where the data was originally taken from \citet{NIST10} and \citet{KP}.
Figure \ref{Spec_Na} shows some examples of observed and synthetic spectra in the \element{Na} doublet region for one GIRAFFE AGB star and one UVES RGB
star. Considering the overall good agreement between the abundances derived from both lines of the \element{Na} doublet, we took the average of the two 
as our final \element{Na} abundance (see Table \ref{Na}). We note that we have not been able to derive a reliable \element{Na} abundance for two out of 
33 AGB stars, owing to their lines approaching saturation. By adopting a solar abundance of $\log\epsilon(\mathrm{Na})_{\odot}=6.24$ \citep{Asplund2009}, 
we derive a mean cluster \element{Na} abundance of [\element{Na}/\element{H}] $=-0.99\,\mathrm{dex}$, with a star-to-star dispersion of 
rms=$0.19\,\mathrm{dex}$. In more detail, we derive [\element{Na}/\element{H}] of $-1.00\,\mathrm{dex}$ (rms=$0.13\,\mathrm{dex}$) and 
[\element{Na}/\element{H}]=$-0.98\,\mathrm{dex}$ (rms=$0.22\,\mathrm{dex}$) for our AGB and RGB samples, respectively. 

\begin{table*}  \small
\caption{Na abundance of our sample stars.
         The complete table is available electronically; we show the first line for guidance.}     
\label{Na}      
\centering      
\begin{tabular}{c c c c c c c c c c}     
\hline\hline      
  Star   & Evol. Ph. & Instrument & $\mathrm{[Na/H]}_{6154}$ & $\mathrm{[Na/H]}_{6154}$ & $\mathrm{[Na/H]}_{6160}$ & $\mathrm{[Na/H]}_{6160}$ & $\mathrm{<[Na/H]>}$ & $\mathrm{<[Na/H]>}$ & $\mathrm{<[Na/\ion{Fe}{i}]>}$     \\ 
           &      &            &   LTE           &   NLTE          &   LTE           &   NLTE          &   LTE     &   NLTE    &   NLTE         \\ 
\hline            
  AGB46601  & AGB &  UVES   &  -0.94  &  -0.99  &  -0.86  &  -0.92  &  -0.90  &  -0.95  &   0.15   \\
$\dots$ & $\dots$ & $\dots$ & $\dots$ & $\dots$ & $\dots$ & $\dots$ & $\dots$ & $\dots$ & $\dots$  \\
\hline         
\end{tabular}
\tablefoot{We show the \element{Na} abundance derived from the \element{Na} line at $6154$\,\AA\ and $6160$\,\AA,\, respectively.
           The three rightmost columns list the final \element{Na} abundance values (i.e. the average of the two line abundances),
           except for those stars that have one of the lines saturated and/or too weak (hence only an upper limit was derived).}
\end{table*}

\subsection{Non-LTE corrections}
\label{section:NLTE}

Since the line formation of neutral iron is sensitive to departures from local thermodynamic equilibrium (LTE) because of its low fraction 
in stellar atmospheres, standard LTE analyses of \ion{Fe}{i} lines tend to underestimate the true iron abundance, while \ion{Fe}{ii} lines 
are not affected (\citealp{Lind2012}, and references therein). However, because the iron abundance derived from \ion{Fe}{i} lines is statistically
more robust owing to the larger number of lines available, we decided to correct our LTE \ion{Fe}{i} abundances for the non-LTE (NLTE) effect 
using the correction grids kindly provided by K. Lind (priv. comm.), the computation of which is documented in \citet{Bergemann2012} 
and \citet{Lind2012}. 

\begin{figure}
    \centering
    \includegraphics[width=0.38\textwidth]{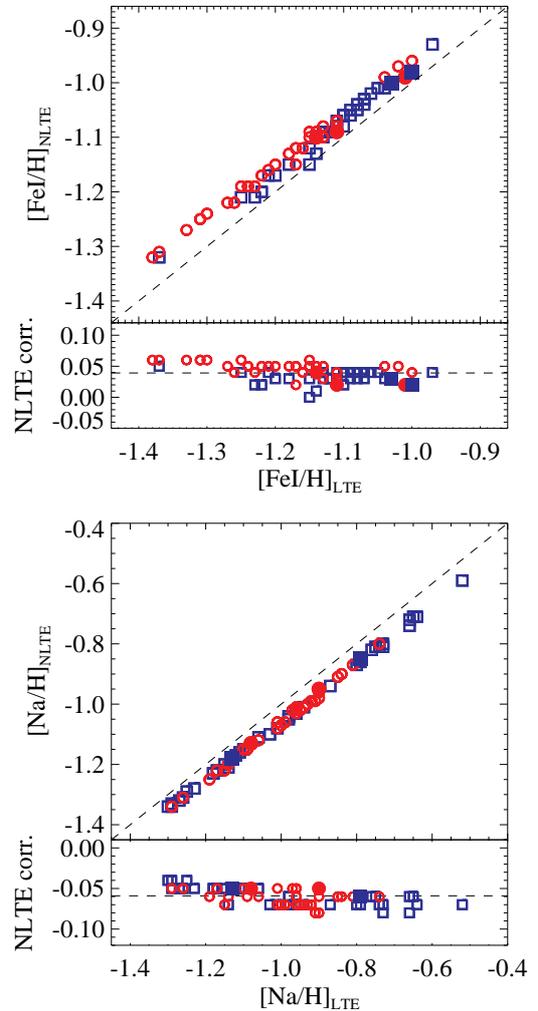}
    \caption{Comparison between NLTE and LTE [\ion{Fe}{i}/\element{H}] (top panel) and [\element{Na}/\element{H}] (bottom panel). 
             Each panel is separated into two plots to show a one-to-one comparison (top) and the distribution of the NLTE 
             correction with the LTE abundance (bottom) where the horizontal dashed line marks the mean NLTE correction. 
             Symbols are the same as in Fig. \ref{CMD}.}
    \label{nlte-lte}
  \end{figure}

The corrections were calculated for each \ion{Fe}{i} line by interpolating the stellar parameters and the EW of the line within the available 
grids. The top panel of Fig. \ref{nlte-lte} shows the overall comparison between NLTE and LTE [\ion{Fe}{i}/\element{H}] and the distribution 
of the NLTE correction. In Table \ref{Abu-nlte}, we list the average LTE and NLTE \ion{Fe}{i} abundances, as well as the NLTE corrections, 
for our AGB and RGB samples respectively. We find that the NLTE correction brings the overall \ion{Fe}{i} metallicity of the cluster to 
[\ion{Fe}{i}/\element{H}]$=-1.11\,\mathrm{dex}$, which is now fully consistent with the value derived from \ion{Fe}{ii} lines. 

Similar to iron, the lines of neutral sodium also form under non-local thermodynamic equilibrium conditions. For \element{Na}, we used the 
correction grids computed by \citet{Lind2011}, which were again applied to each LTE \element{Na} line abundance, selecting the exact stellar 
parameters of the star under investigation. As our final \element{Na} abundance per star, we took the average value of the NLTE \element{Na} 
abundances derived from each line of the doublet (see Table \ref{Na}). The comparison between the NLTE and LTE [\element{Na}/\element{H}]
abundance ratios and the NLTE correction distribution derived for our sample is shown in the bottom panel of Fig. \ref{nlte-lte}. 
The NLTE correction shifts the \element{Na} abundance downwards systematically. The LTE and NLTE [\element{Na}/\element{H}], together with 
the NLTE corrections of \element{Na} abundance, are also listed in Table \ref{Abu-nlte}. 

\begin{table*}  \small
\caption{LTE and NLTE mean abundances of \ion{Fe}{i} and \element{Na} of our sample.}        
\label{Abu-nlte}      
\centering            
\begin{tabular}{c c c c c c c c c c c}     
\hline\hline        
  Sample  &  [\ion{Fe}{i}/\element{H}]$_\mathrm{LTE}$ &  rms  & [\ion{Fe}{i}/\element{H}]$_\mathrm{NLTE}$ & rms & NLTEcorr(\ion{Fe}{i}) & [\element{Na}/\element{H}]$_\mathrm{LTE}$ &  rms  & [\element{Na}/\element{H}]$_\mathrm{NLTE}$ & rms &  NLTEcorr(\element{Na})  \\
          &     ($\mathrm{dex}$)                & ($\mathrm{dex}$) &     ($\mathrm{dex}$)         & ($\mathrm{dex}$) & ($\mathrm{dex}$) &     ($\mathrm{dex}$)                & ($\mathrm{dex}$) &     ($\mathrm{dex}$)         & ($\mathrm{dex}$) & ($\mathrm{dex}$)     \\
\hline     
  Total   &  -1.15   &  0.09   &  -1.11   &  0.09   &  0.039   &  -0.99   &  0.19   &  -1.05   &  0.18   &  -0.060       \\
   AGB    &  -1.19   &  0.10   &  -1.14   &  0.09   &  0.047   &  -1.00   &  0.13   &  -1.06   &  0.13   &  -0.062       \\
   RGB    &  -1.12   &  0.07   &  -1.08   &  0.07   &  0.033   &  -0.98   &  0.22   &  -1.04   &  0.21   &  -0.059       \\
\hline      
\end{tabular}
\end{table*}

\subsection{Error analysis}
\label{section:Error}

Before discussing our abundance results, we need to estimate the error on the derived abundances. Several sources contribute to the 
final uncertainty, both of random and systematic nature. 

The random measurement uncertainty on the derived abundances can generally be estimated by $\sigma/\sqrt{\mathrm{N}}$, where $\sigma$ 
is the line-to-line dispersion and $\mathrm{N}$ is the number of lines measured. 
Here we note that the UVES and GIRAFFE samples have slightly different random uncertainties owing to their different spectral resolutions and 
$\lambda-$coverages. Considering the limited number of \element{Na} and \ion{Fe}{ii} lines present in our spectra, the application of the above 
formula is likely less accurate. To compensate for the low number statistics of the \element{Na} and \ion{Fe}{ii} indicators, we thus applied a 
correction to the value of $\sigma/\sqrt{\mathrm{N}}$ according to a t-distribution requiring a 1$\sigma$ confidence level.

The systematic measurement uncertainty is usually estimated by evaluating the effect on the derived abundances of varying stellar parameters 
and EWs (or other key parameters of the analysis) by their associated errors, keeping in mind that the stellar parameters are mutually dependent. 
For this purpose, we selected six representative stars of our samples (GIRAFFE: cool/hot, AGB/RGB, one each; UVES: AGB/RGB, both cool since the 
UVES sample only has cool stars). 

To estimate the systematic influence of the derived stellar parameters, we changed each of the input values ($T_\mathrm{eff}$, $\log g$, 
[\element{M}/\element{H}], and $\xi_\mathrm{t}$) in turn, by the amounts corresponding to the uncertainties we derived for each of them 
($\pm70\,\mathrm{K}$, $\pm0.05\,\mathrm{dex}$, $\pm0.1\,\mathrm{dex}$, and $\pm0.1\,\mathrm{km\,s}^{-1}$ respectively).  
When changing one parameter, we also iteratively updated all other parameters (see Sect. \ref{section:Metallicity}). Considering the very 
similar dependences found for all GIRAFFE and for all UVES stars, Table \ref{error_test} only gives the `sample'-average values. 

For the uncertainty on the individual EW measurements, we adopted the errors estimated by DAOSPEC ($\mathrm{e}_\mathrm{daospec}\mathrm{(EW)}$),
which are derived during the least-square fit of a given line. These uncertainties correspond to an average variation 
of $\pm0.04\,\mathrm{dex}$ in [\ion{Fe}{i}/\element{H}] and $\pm0.06\,\mathrm{dex}$ in [\ion{Fe}{ii}/\element{H}] for the GIRAFFE sample, 
and $\pm0.02\,\mathrm{dex}$ in [\ion{Fe}{i}/\element{H}] and $\pm0.03\,\mathrm{dex}$ in [\ion{Fe}{ii}/\element{H}] for the UVES sample. 

For the uncertainties in the \element{Na} abundances, which have been determined via spectrum synthesis, we started by shifting the spectra continuum
by $\pm0.5\%$ and found an average variation of $\pm0.06\,\mathrm{dex}$ in [\element{Na}/\element{H}] for the GIRAFFE sample and $\pm0.02\,\mathrm{dex}$
for the UVES sample. Another possible source of uncertainty is the choice of the atomic physics. Our analysis made use of the atomic parameters of 
the \element{Na} doublet as reported in the VALD3 database, i.e. $\log(gf)=-1.547$ for the \element{Na} line at $6154$\,\AA\ and $-1.246$ for the one 
at $6160$\,\AA. No uncertainty is reported for either value, even in the original sources. However, from fitting the solar spectrum with an NLTE model
atmosphere, \citet{Gehren2004} derived a slightly different pair of $\log(gf)$ values, $-1.57$ and $-1.28$ for the $6154$\,\AA\ and $6160$\,\AA\ \element{Na} 
lines, respectively. If we now round off this small difference ($0.03$) to $\pm 0.05$ as our uncertainty on the oscillator strength, we find 
a $\mp0.05\,\mathrm{dex}$ dependence of the derived [\element{Na}/\element{H}] ratio for all stars. 

\begin{table}  \footnotesize
\caption{Sensitivities of \element{Fe} and \element{Na} abundances to the variations in stellar parameters.}     
\label{error_test}      
\centering              
\begin{tabular}{c c c}      
\hline\hline         
                         & GIRAFFE sample & UVES sample     \\
\hline            
$\Delta T_\mathrm{eff} = \pm 70\,\mathrm{K}$  &  &    \\
 $\Delta$[\ion{Fe}{i}/H]  &  $\pm 0.05$  &  $\pm 0.03$     \\
 $\Delta$[\ion{Fe}{ii}/H] &  $\mp 0.06$  &  $\mp 0.08$     \\
 $\Delta$[Na/H]           &  $\pm 0.06$  &  $\pm 0.08$     \\
\hline               
$\Delta \log g = \pm 0.05\,\mathrm{dex}$  &  &    \\
 $\Delta$[\ion{Fe}{i}/H]  &  $\pm 0.01$  &  $\pm 0.01$     \\
 $\Delta$[\ion{Fe}{ii}/H] &  $\pm 0.03$  &  $\pm 0.02$     \\
 $\Delta$[Na/H]           &  $\pm 0.01$  &  $\pm 0.01$     \\
\hline               
$\Delta \xi_\mathrm{t} = \pm 0.1\,\mathrm{km\,s}^{-1}$  &  &   \\
 $\Delta$[\ion{Fe}{i}/H]  &  $\mp 0.04$  &  $\mp 0.04$     \\
 $\Delta$[\ion{Fe}{ii}/H] &  $\mp 0.02$  &  $\mp 0.03$     \\
 $\Delta$[Na/H]           &  $\pm 0.01$  &  $\pm 0.02$     \\
\hline               
$\Delta \mathrm{[M/H]} = \pm 0.1\,\mathrm{dex}$  &  &    \\
 $\Delta$[\ion{Fe}{i}/H]  &  $\pm 0.01$  &  $\pm 0.01$     \\
 $\Delta$[\ion{Fe}{ii}/H] &  $\pm 0.04$  &  $\pm 0.04$     \\
 $\Delta$[Na/H]           &  $\mp 0.01$  &  $\mp 0.01$     \\
\hline 
\end{tabular} 
\end{table}

Table \ref{Error} summarises our complete error analysis providing the overall random/systematic/total uncertainties as derived separately 
for our GIRAFFE and UVES samples by taking the square root of the quadratic sum of the errors associated to all factors, so far discussed. 
Because the four stellar parameters are mutually dependent, the systematic dependences (and, in turn, the total errors) may be too conservative 
but they help compensate, at least in part, for those random uncertainties that it is not possible to properly account for. 
As already mentioned, the difference between the values derived for the two samples (GIRAFFE and UVES) is a consequence of their different 
spectral resolutions, wavelength coverages, and temperature ranges (e.g. the cool stars observed with UVES only overlap with the coolest stars 
observed with GIRAFFE). 

\begin{table}  \footnotesize
\caption{Uncertainties on the \element{Fe} and \element{Na} abundances.}     
\label{Error}      
\centering         
\begin{tabular}{c c c}      
\hline\hline      
                          & GIRAFFE sample &  UVES sample      \\
\hline    
Random                    & &   \\
$\Delta$[\ion{Fe}{i}/H]   &  $\pm 0.02$   &  $\pm 0.01$        \\
$\Delta$[\ion{Fe}{ii}/H]  &  $\pm 0.05$   &  $\pm 0.02$        \\
$\Delta$[Na/H]            &  $\pm 0.04$   &  $\pm 0.04$        \\
$\Delta$[Na/\ion{Fe}{i}]  &  $\pm 0.04$   &  $\pm 0.04$        \\
$\Delta$[Na/\ion{Fe}{ii}] &  $\pm 0.06$   &  $\pm 0.04$        \\
\hline         
Systematic                & &   \\
$\Delta$[\ion{Fe}{i}/H]   &  $\pm 0.08$   &  $\pm 0.06$        \\
$\Delta$[\ion{Fe}{ii}/H]  &  $\pm 0.10$   &  $\pm 0.10$        \\
$\Delta$[Na/H]            &  $\pm 0.10$   &  $\pm 0.10$        \\
$\Delta$[Na/\ion{Fe}{i}]  &  $\pm 0.09$   &  $\pm 0.09$        \\
$\Delta$[Na/\ion{Fe}{ii}] &  $\pm 0.15$   &  $\pm 0.18$        \\
\hline 
Total                     & &   \\
$\Delta$[\ion{Fe}{i}/H]   &  $\pm 0.08$   &  $\pm 0.06$        \\
$\Delta$[\ion{Fe}{ii}/H]  &  $\pm 0.11$   &  $\pm 0.10$        \\
$\Delta$[Na/H]            &  $\pm 0.11$   &  $\pm 0.11$        \\
$\Delta$[Na/\ion{Fe}{i}]  &  $\pm 0.10$   &  $\pm 0.10$        \\
$\Delta$[Na/\ion{Fe}{ii}] &  $\pm 0.16$   &  $\pm 0.18$        \\
\hline 
\end{tabular} 
\end{table}

In Fig. \ref{Spec_Na} we show the combined effect of random and fitting errors in the \element{Na} abundance ($\pm0.07\,\mathrm{dex}$ for 
GIRAFFE star and $\pm0.05\,\mathrm{dex}$ for UVES star) on the synthesised \element{Na} line profiles.

\subsection{Stars in common with \citet{Carretta2006} work on NGC\,2808}
\label{section:comp_Carretta2006}

\begin{figure}
    \centering
    \includegraphics[width=0.44\textwidth]{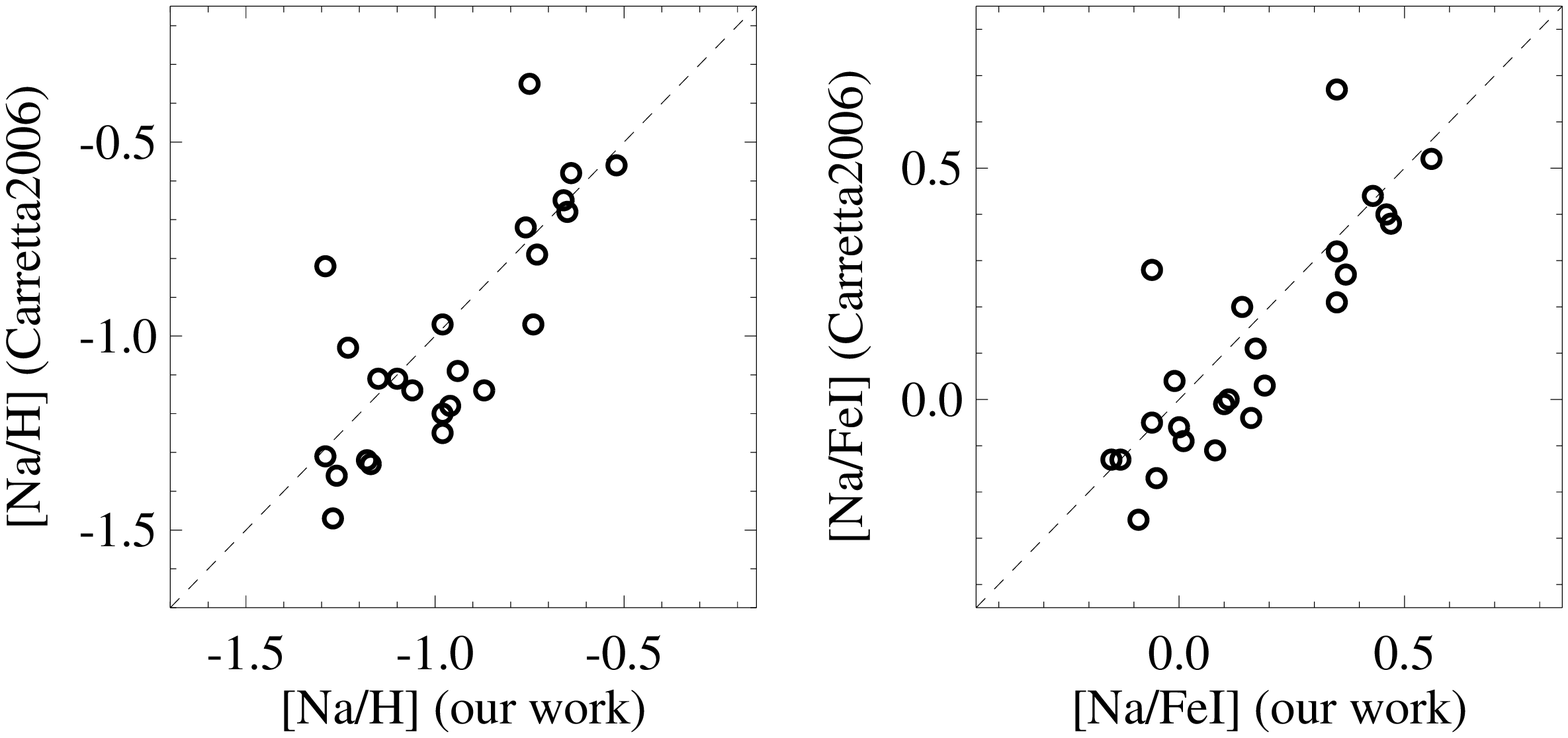}
    \caption{Comparison of [\element{Na}/\element{H}] and [\element{Na}/\ion{Fe}{i}] ratio of the common RGB stars 
             between our work and \citet{Carretta2006}. Here only LTE abundances are considered.}
    \label{common-stars0}
  \end{figure}

By cross-matching target coordinates (within an angular distance of $<0.3''$), we were able to identify $24$ RGB stars in common with the sample 
of \citet{Carretta2006}. On average, we find a good agreement on most stellar parameters and abundances. The differences between the two analyses 
(here, always reported as ``this work $-$ \citet{Carretta2006}'') are negligible in the stellar parameters, \ion{Fe}{i} and Na abundance
values\footnote{$\Delta T_\mathrm{eff}=-13\pm25\,\mathrm{K}$, $\Delta\log g=+0.02\pm0.01$, $\Delta\xi_\mathrm{t}=-0.04\pm0.18\,\mathrm{km\,s}^{-1}$, 
$\Delta[\ion{Fe}{i}/\element{H}]_\mathrm{LTE}=0.00\pm0.08\,\mathrm{dex}$, $\Delta[\element{Na}/\element{H}]_\mathrm{LTE}=+0.04\pm0.19\,\mathrm{dex,}$
and $\Delta[\element{Na}/\ion{Fe}{i}]_\mathrm{LTE}=+0.04\pm0.13\,\mathrm{dex}$}, while for \ion{Fe}{ii} amount to $+0.09\pm0.14\,\mathrm{dex}$. 
As an example, Fig. \ref{common-stars0} shows the comparisons between us and them for both the [\element{Na}/\element{H}]$_\mathrm{LTE}$ and 
[\element{Na}/\ion{Fe}{i}]$_\mathrm{LTE}$ ratios.

Applying the sensitivities of \element{Fe} and \element{Na} abundances to the stellar parameters as reported in Table \ref{error_test}, we can 
actually account for some of the abundance differences listed above. Small offsets ($\sim+$0.05\,dex) in \ion{Fe}{ii} and \element{Na} remain, 
but they are well within the associated error-bars. We are thus quite confident that our derived abundances are overall in good agreement with 
those derived by \citet{Carretta2006}.
 
The remaining, albeit small, differences could also be due to the different input values and/or methodologies employed by us and by \citet{Carretta2006} 
(different photometry, $T_\mathrm{eff}-$colour calibrations, abundance derivations). Because the final goal of our project is the accurate determination 
of \element{Na} abundances, we think it is important to track these down in more detail. As far as the temperature is concerned, we find a mean 
$\Delta T_\mathrm{eff}=+$44\,K because of the different $T_\mathrm{eff}-$colour calibrations used, but we note that we are able to reproduce 
\citet{Carretta2006} $T_\mathrm{eff}$ and $\log g$ values when we adopt their photometry and colour-temperature calibration. 

\begin{figure}
    \centering
    \includegraphics[width=0.44\textwidth]{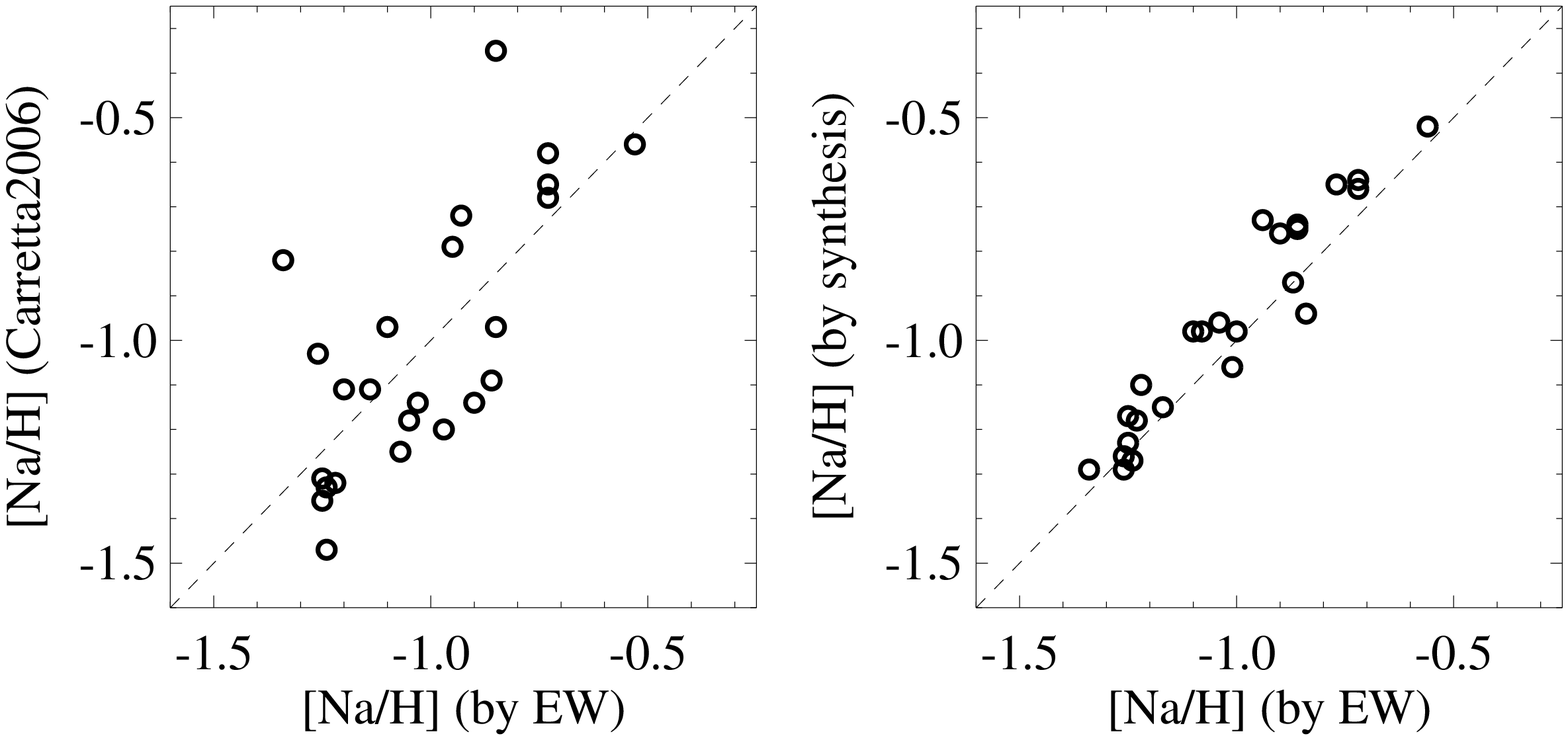}
    \caption{Comparison of [\element{Na}/\element{H}], as a result of the method test considering the common RGB stars with \citet{Carretta2006}. 
             The left panel compares the abundances re-derived by us based on the EWs of the 6154$-$60\AA\ doublet and the set of  stellar parameters 
             derived according to \citet{Carretta2006} prescriptions with those derived by \citet{Carretta2006}. 
             The right panel shows the good agreement between the spectrum-synthesis-based abundances and the (6154$-$60\AA\ doublet) EW-based values, 
             based on our own set of stellar parameters. 
             Here only LTE abundances are considered.}
    \label{common-stars}
  \end{figure}

For the \element{Na} abundances, the comparison is between spectrum synthesis (ours) and EW-based abundances (\citealp{Carretta2006}), but it is 
more cumbersome because we used only the 6154$-$6160\AA\ doublet, while \citet{Carretta2006} derived their \element{Na} abundances from the EW of 
different \element{Na} doublets, depending on the available spectra. Moreover, the 6160\AA\ line appears to be severely blended with a neighbouring 
Ca line on its red wing for several of our RGB stars, making the EW method less reliable. Nonetheless, we find good agreement on the mean values 
when we derive the \element{Na} abundances from the EW of the 6154$-$6160\AA\ doublet and with the set of stellar parameters derived according 
to \citet{Carretta2006} prescriptions ($\Delta\mathrm{[Na/H]} \sim -0.01\pm0.21\,\mathrm{dex}$, see left panel of Fig. \ref{common-stars}), although 
the amount of scatter persists, mostly as a result of the errors associated with the measurement of the equivalent widths. We also find that, based on 
our determination of stellar parameters, the spectrum-synthesis-based \element{Na} values are, on average, only $\sim0.05\pm0.07\,\mathrm{dex}$
higher than our EW-based values (see. Fig. \ref{common-stars}, right panel).

Overall, we can then conclude that the different methods explored so far for the derivation of stellar parameters and/or abundances lead to small
systematic differences (within the associated errors in our case), while the errors associated with the EWs/line-fitting measurements are mostly 
responsible for the dispersions around these values. To complete our diagnostic tests, we have also evaluated the effects of deriving the stellar
parameters spectroscopically on our final \element{Na}. The results of this test are summarised in the Appendix. 

\section{Observed Na distribution along the RGB and AGB in NGC\,2808}

\subsection{[Na/H] versus [Na/Fe]}

  \begin{figure*}
    \centering
    \includegraphics[width=0.95\textwidth]{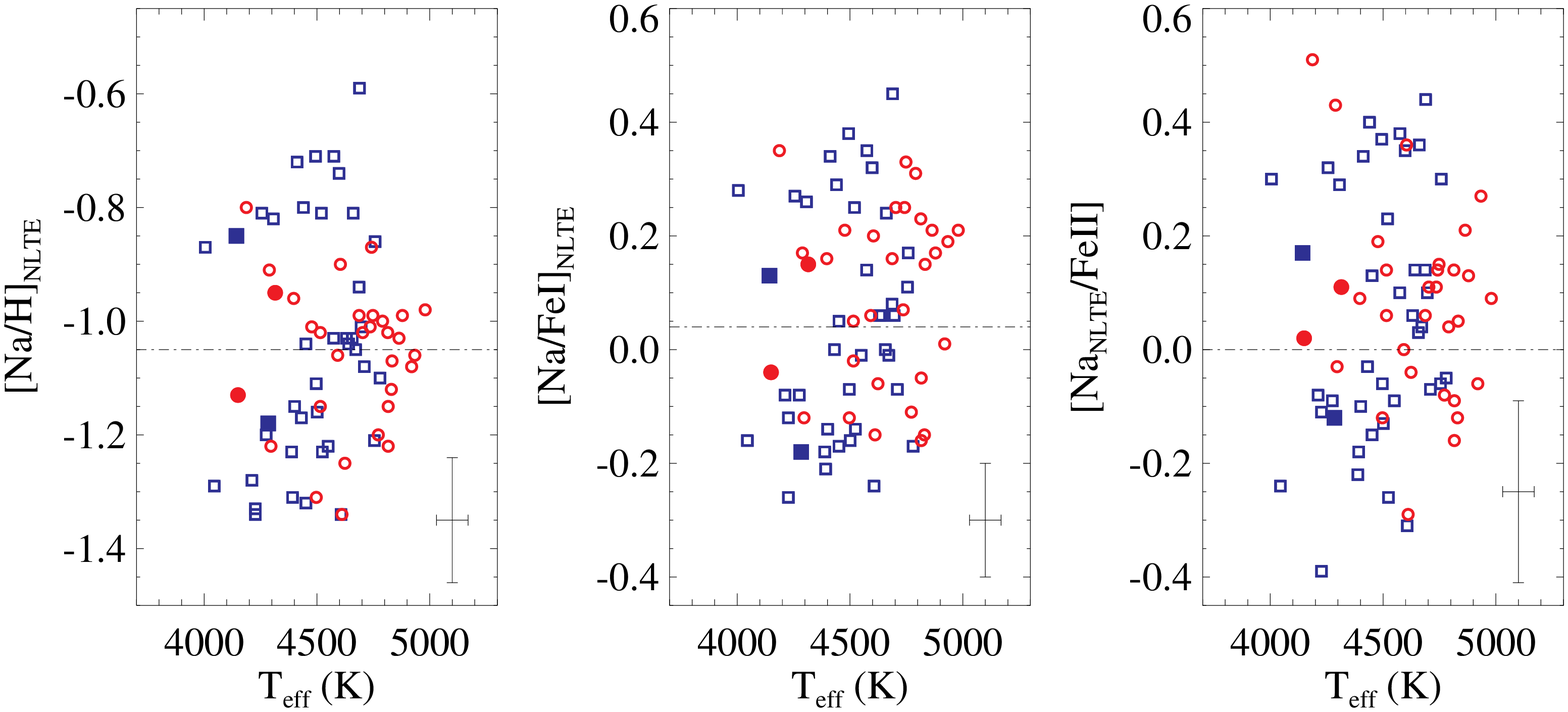}
    \caption{Abundance distributions of our complete (AGB $+$ RGB) sample. 
             {\it Left:} [\element{Na}/\element{H}]$_\mathrm{NLTE}-T_\mathrm{eff}$; 
             {\it middle:} [\element{Na}/\ion{Fe}{i}]$_\mathrm{NLTE}-T_\mathrm{eff}$; 
             {\it right:} [\element{Na}$_\mathrm{NLTE}$/\ion{Fe}{ii}]$-T_\mathrm{eff}$. 
             Symbols are the same as in Fig. \ref{CMD}. 
             The horizontal dash-dotted lines mark the critical values distinguishing the 1P and 2P stars 
             following \citet{Carretta2009a} criteria (see the text). 
             The typical error bars are shown at the right-bottom corner of each panel (considering the similarity of 
             the errors of the GIRAFFE and UVES samples and the clarity of the figure, we only show the error bars for 
             the GIRAFFE sample which is the largest sample here).}
    \label{Na_distr}
  \end{figure*}

By comparing the errors in our derived metallicities ($\sigma_\mathrm{FeI}=0.08\,\mathrm{dex}$) and the star-to-star dispersions 
($\sigma_\mathrm{FeI,obs}=0.09\,\mathrm{dex}$), we find the intrinsic spread in [\element{Fe}/\element{H}] of NGC\,2808 to be within 
$\sim0.05\,\mathrm{dex}$, when considering only the GIRAFFE sample (representing the large majority of our dataset). The stellar population 
in NGC\,2808 can thus be considered homogeneous in its \element{Fe} content within a few hundredths of $\mathrm{dex}$, as also pointed out 
by \citet{Carretta2006}. 

Figure \ref{Na_distr} shows these effects on the observed abundance patterns, where we present our final \element{Na} abundance distributions for 
AGB and RGB stars in NGC\,2808 as [\element{Na}/\element{H}], [\element{Na}/\ion{Fe}{i}], and [\element{Na}/\ion{Fe}{ii}] versus $T_\mathrm{eff}$. 
Overall, the differences among the three panels are minimal. However, because of the extra uncertainties associated with \ion{Fe}{i} (due to the 
NLTE corrections) and \ion{Fe}{ii} (due to the paucity of lines), we chose to use the [\element{Na}/\element{H}] ratio to discuss our derived \element{Na} 
abundance distributions along the RGB and AGB.

\subsection{[Na/H] distribution}

A quick inspection of Fig. \ref{Na_distr} (left panel) shows that the AGB and RGB samples in NGC\,2808 have similar \element{Na} abundance ranges. 
A two-sided Kolmogorov-Smirnov (K-S) test confirms this: the K-S statistic D (0.268) and the p-value (0.137) derived for the [\element{Na}/\element{H}] 
distribution both indicate that there is only weak evidence to reject the null hypothesis, i.e. the two samples have the same distribution. Examining 
in more detail the dispersions ($\sigma$) and the interquartile range (IQR) values of [\element{Na}/\element{H}] of the AGB and RGB samples 
($\sigma_\mathrm{AGB}=0.13$ and $\sigma_\mathrm{RGB}=0.21$, IQR$_\mathrm{AGB}=0.16$ and IQR$_\mathrm{RGB}=0.39$), we find that the RGB sample is more 
evenly spread across the entire \element{Na} abundance range, while the AGB stars tend to be more concentrated. We also find that the maximum 
[\element{Na}/\element{H}] value derived for the AGB sample is 0.21\,dex lower than the one derived for the RGB sample. These can be seen in Fig. 
\ref{NaH_hist} where we show the histograms and cumulative distributions of [\element{Na}/\element{H}] for both the AGB and the RGB samples. 

The conclusions are relatively similar when we turn to [\element{Na}/\ion{Fe}{i}] and [\element{Na}/\ion{Fe}{ii}] (Fig. \ref{Na_distr}, middle and right). 
For [\element{Na}/\ion{Fe}{i}] we obtain (D, p-value)$=$(0.248, 0.199), $\sigma_\mathrm{AGB,RGB}=(0.16,0.20)$, IQR$_\mathrm{AGB,RGB}=(0.26,0.39)$; 
whereas for [\element{Na}/\ion{Fe}{ii}] we derive (D, p-value)$=$(0.224, 0.304), $\sigma_\mathrm{AGB,RGB}=(0.17,0.23)$ and IQR$_\mathrm{AGB,RGB}=(0.18,0.41)$. 

\subsection{Fraction of 1P and 2P stars along the RGB and the AGB}

\begin{figure}
    \centering
    \includegraphics[width=0.37\textwidth]{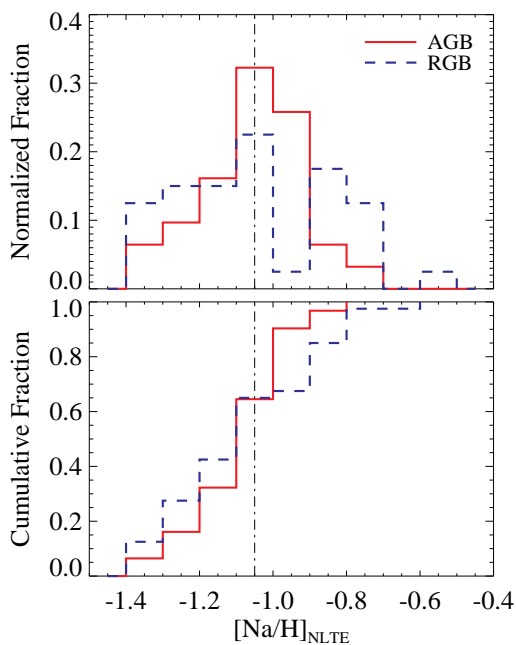}
    \caption{Histograms of [\element{Na}/\element{H}] (top) and cumulative distribution functions (bottom) of our AGB and RGB sample in NGC\,2808. 
             The black vertical dash-dotted line separates 1P and 2P stars using \citet{Carretta2009a} criteria (see text).} 
    \label{NaH_hist}
  \end{figure}

To estimate the relative fraction of 1P vs 2P stars in our AGB and RGB samples in terms of \element{Na} enrichment, we follow \citet{Carretta2009a} 
who distinguishes 1P and 2P stars by identifying those stars that have, respectively, [\element{Na}/\element{Fe}] below and above 
[\element{Na}/\element{Fe}]$_\mathrm{cri}=$ [\element{Na}/\element{Fe}]$_\mathrm{min}+0.3\,\mathrm{dex}$, 
where [\element{Na}/\element{Fe}]$_\mathrm{min}$ is the minimum value of [\element{Na}/\element{Fe}] derived for the entire sample. 
We extend this criteria to the absolute abundance and take 
[\element{Na}/\element{H}]$_\mathrm{cri}=$ [\element{Na}/\element{H}]$_\mathrm{min}+0.3\,\mathrm{dex}$ 
as the actual reference value to separate 1P and 2P stars. 
The black dash-dotted lines in Figs. \ref{Na_distr} and \ref{NaH_hist} mark this critical value in each distribution, roughly separating 
\element{Na}-poor 1P from \element{Na}-rich 2P stars. It clearly appears that NGC\,2808 does host \element{Na}-rich 2P AGB stars. 
When considering the [\element{Na}/\element{H}] distribution and using the \element{Na} cut-off proposed above, the ratio of 1P to 2P stars 
is 45:55 and 48:52 in the AGB and RGB samples, respectively.  
As a sanity check, even when [\element{Na}/\ion{Fe}{i}] or [\element{Na}/\ion{Fe}{ii}] is considered, one still finds more \element{Na}-rich than 
\element{Na}-poor AGB stars, while these two populations are comparable in number along the RGB. Therefore, in our sample, 2P AGB stars are more 
numerous than 2P RGB stars when compared to their respective 1P counterparts, although they do not reach the same maximum \element{Na} abundance as 
found on the RGB. Furthermore, as far as the RGB stars are concerned, our results agree well with \citet{Carretta2009a} who analysed a much larger 
RGB star sample (98 stars) and found a 1P:2P stars ratio of 50:50, and even better with \citet{Carretta2015ApJ} who reported a similar ratio 
(1P:2P$=$46:54) from the \element{O}$-$\element{Na} anti-correlation of 140 RGB stars. 

\section{Discussion}
\label{section:discussion}

After the finding by \citet{Campbell2013} that NGC\,6752 lacks \element{Na}-rich AGB stars, recent studies have revealed a complex chemical picture 
of AGB stars in Galactic GCs. While another cluster (M\,62) was found to be devoid of \element{Na}-rich 2P AGB stars, other GCs (NGC\,104, M\,13, M\,5, 
M\,3, and M\,2) do have 2P AGB stars (see references in Sect. \ref{section:introduction}). This work on NGC\,2808 adds one more cluster to the latter group. 
This is particularly interesting, as it provides important constraints on our understanding of the formation and evolution of GCs and of their stellar 
populations. 

It is now largely accepted that GCs suffered from self-enrichment during their early evolution, and that 2P stars formed out of the \element{Na}-rich, 
\element{O}-poor ashes of hydrogen burning at high temperature ejected by more massive 1P stars and diluted with interstellar gas (e.g. 
\citealp{PrantzosCharbonnel2006, Prantzos2007}). Among the most commonly-invoked sources of H-burning ashes, one finds 
fast-rotating massive stars (FRMS, with initial masses above $\sim25\,\mathrm{M_{\sun}}$; \citealp{MaederMeynet2006, PrantzosCharbonnel2006, 
Decressin2007a, Decressin2007b, Krause2013}) and massive AGB stars (with initial masses of $\sim6-11\,\mathrm{M_{\sun}}$; 
\citealp{Ventura2001, DErcole2010, Ventura2011, Ventura2013}). The role of other possible polluters has also been explored: massive stars in close 
binaries ($10-20\,\mathrm{M_{\sun}}$; \citealp{deMink2009, Izzard2013}), FRMS paired with AGB stars \citep{Sills2010} or with high-mass interactive 
binaries \citep{Bastian2013, Cassisi2014a}, and supermassive stars ($\sim10^{4}\,\mathrm{M_{\sun}}$; \citealp{Denissenkov2014}). 

The various scenarios for GC self-enrichment differ on many aspects. The most commonly-invoked ones, in particular, make very different predictions 
for the coupling between \element{Na} and \element{He} enrichment in the initial composition of 2P stars. In the AGB scenario, on the one hand, 
all 2P stars spanning a large range in \element{Na} are expected to be born with very similar \element{He} (maximum $\sim$ 0.36 $-$ 0.38 in mass fraction). 
This is due to the fact that \element{He} enrichment in the envelope of the intermediate-mass stellar polluters results from the second dredge-up 
on the early-AGB before the TP-AGB, where hot bottom-burning might affect the abundances of \element{Na} in the polluters yields 
\citep[e.g.][]{Forestini1997,Ventura2013,Doherty2014}. Therefore within this framework one is expecting to find the same proportion of 1P and 2P stars 
in the various regions of the GC CMDs, at odds with the observations in NGC\,6752. 

In the FRMS scenario, on the other hand, the \element{Na} and \element{He} enrichment of 2P stars at birth are correlated since both chemical elements 
result from simultaneous H-burning in fast-rotating massive main-sequence stars (\citealt{Decressin2007b,Decressin2007a,Chantereau2015}).
This has important consequences on the way 1P and 2P stars populate the various sequences of the CMDs.
Indeed, owing to the impact of the initial \element{He} content of stars on their evolution paths and lifetimes, 2P stars that are born with a chemical 
composition above an initial \element{He} and \element{Na} abundances cut-off do not climb the AGB and evolve directly towards the white dwarf stage after 
central \element{He} burning \citep{Chantereau2015,Chantereau2016}. 
Therefore, the coupling between \element{He} and \element{Na} enrichment in the initial composition of 2P stars predicted by the FRMS scenario is 
expected to lead an evolution of the \element{Na} dispersion between the RGB and AGB in individual GCs, in proportions that depend on their age and 
metallicity \citep{CCWC2016}. 
The corresponding theoretical predictions for an old and relatively metal-poor GC like NGC\,6752 ([\element{Fe}/\element{H}]$=-1.54$, \citealt{Campbell2013}; 
age between $\sim12.5\pm0.25\,\mathrm{Gyr}$ and $13.4\pm1.1\,\mathrm{Gyr}$ according to \citealt{VandenBerg2013} and \citealt{Gratton2003} respectively) 
are in good agreement with the lack of \element{Na}-rich AGBs in this cluster, although the precise \element{Na}-cut-off on the AGB is expected to depend 
on the assumed RGB mass-loss rate \citep{Charbonnel2013}. 

\citet{CCWC2016} then showed that within the FRMS scenario the maximum \element{Na} content expected for 2P stars on the AGB is a function of both 
the metallicity and the age of GCs. Namely, at a given [\element{Fe}/\element{H}], younger clusters are expected to host AGB stars exhibiting a larger 
\element{Na} spread than older clusters; and, at a given age, higher \element{Na} dispersion along the AGB is predicted in metal-poor GCs than in the 
metal-rich ones. NGC\,2808 is both younger ($11.00\pm0.38\,\mathrm{Gyr}$ by \citealt{VandenBerg2013} and $10.9\pm0.7\,\mathrm{Gyr}$ by 
\citealt{Massari2016}) and more-metal rich than NGC\,6752 and it lies in the domain where RGB and AGB stars are expected to present very similar 
dispersions, as predicted within the FRMS framework \citep{CCWC2016}. Therefore, to first order, the present observations seem to agree well with the 
theoretically predicted trends. 

However, additional parameters may play a role in inducing cluster-to-cluster variations, as already suggested by the spectroscopic and photometric 
diversity of these complex stellar systems. In particular, mass loss in the earlier phases of stellar evolution (RGB) has been shown to impact the 
\element{Na} cut on the AGB; the higher the mass loss, the stronger the expected differences with age and metallicity between RGB and AGB stars 
(\citealt{CCWC2016}; see also \citealp{Cassisi2014b}). In addition, the maximum mass of the FRMS polluters might change from cluster to cluster, which 
should affect their yields and therefore the shape and the extent of the \element{He}-\element{Na} correlation for 2P stars. It is therefore fundamental 
to gather additional data in an homegeneous way for GCs spanning a large range in age, metal content, and general properties (mass, compactness, etc) 
to better constrain the self-enrichment scenarios. Work is in progress in this direction and we will present new \element{Na} abundance determinations 
in three other GCs in the second paper of this series. 

\section{Summary}

The current sample of GC AGB stars available in the literature with accurately derived \element{Na} abundances remains rather limited. 
To increase the number statistics and to further characterise the presence and nature of \element{Na}-rich stars on the AGB, we observed and analysed 
a new sample of 33 AGB and 40 RGB stars in the Galactic GC NGC\,2808. 

We applied standard analytical methods to derive the stellar parameters and metallicities of the sample. Effective temperatures and stellar gravities were 
determined photometrically, while the metallicity was derived via equivalent widths of (several) \ion{Fe}{i} and (fewer) \ion{Fe}{ii} absorption lines. 
Since the \ion{Fe}{i} abundance is affected by the NLTE effect, we applied the NLTE correction, which increased the mean [\ion{Fe}{i}/\element{H}] by 
$\sim0.04\,\mathrm{dex}$. Here the [\ion{Fe}{i}/\element{H}]$_\mathrm{NLTE}$ and [\ion{Fe}{ii}/\element{H}] agree well. We thus derived a mean metallicity 
of NGC\,2808 of $-1.11\pm0.08\,\mathrm{dex}$. 

We tested the influence on the final \element{Na} abundance of adopting photometric vs. spectroscopic methods in the derivation of the stellar parameters. 
Our test shows that this effect is not significant compared to the various sources of uncertainty, especially when discussing [\element{Na}/\element{H}] 
and/or [\element{Na}/\ion{Fe}{i}] abundance ratios. 

Sodium abundances were derived for 30 AGB and 40 RGB stars. From our results, AGB and RGB stars in NGC\,2808 have comparable overall \element{Na} 
distributions. By examining the dispersion, the interquartile range and the maximum value of the [\element{Na}/\element{H}] ratio determined in the 
AGB and RGB samples, the former appears more concentrated than the latter, in terms of \element{Na} abundance ratios. 

Following the same criteria as proposed by \citet{Carretta2009a} to separate 1P and 2P GC stars, we derive 1P/2P star ratios of 45:55 (AGB sample) and
48:52 (RGB sample), when the [\element{Na}/\element{H}] abundance ratio is considered. This result shows that NGC\,2808 has an asymptotic giant branch that 
is populated by relatively larger numbers of \element{Na}-rich stars than \element{Na}-poor ones, while the two groups are of comparable size on the cluster 
RGB. This work thus adds another slightly metal-poor GC, NGC\,2808, to the group of galactic globular clusters that have \element{Na}-rich 2P AGB stars. 

When compared to theoretical models, our finding are well accounted for by the FRMS scenario, without being in contradiction with earlier results 
from, e.g. \citet{Campbell2013}. It seems thus important to better quantify the dependences of the \element{Na} distributions from cluster metallicity 
and age. A detailed comparison of \element{Na} abundances on the asymptotic giant branch of several GCs and theoretical predictions will be presented 
in a forthcoming paper (Wang et al. in preparation).

\begin{acknowledgements}

YW acknowledges the support from the European Southern Observatory, via its ESO Studentship programme. 
This work was partly funded by the National Natural Science Foundation of China under grants 1233004 and 11390371, 
as well as the Strategic Priority Research Program The Emergence of Cosmological Structures of the Chinese Academy
of Sciences, Grant No. XDB09000000. 
CC and WC acknowledge support from the Swiss National Science Foundation (FNS) for the project 200020-159543 
Multiple stellar populations in massive star clusters - Formation, evolution, dynamics, impact on galactic evolution. 
We are indebted to Peter Stetson for kindly providing us with accurate Johnson-Morgan photometry. 
We thank Karin Lind for useful discussions and for giving us access to her NLTE correction grids. 
We thank the International Space Science Institute (ISSI, Bern, CH) for welcoming the activities of ISSI Team 271 
Massive star clusters across the Hubble Time (2013 - 2016). 
This work has made use of the VALD database, operated at Uppsala University, the Institute of Astronomy RAS in Moscow, 
and the University of Vienna. 
We thank the anonymous referee for the detailed comments and useful suggestions to improve the paper. 

\end{acknowledgements}



\begin{thebibliography}{104}
\expandafter\ifx\csname natexlab\endcsname\relax\def\natexlab#1{#1}\fi

\bibitem[{{Alonso} {et~al.}(1999){Alonso}, {Arribas}, \&
  {Mart{\'{\i}}nez-Roger}}]{Alonso1999}
{Alonso}, A., {Arribas}, S., \& {Mart{\'{\i}}nez-Roger}, C. 1999, \aaps, 140,
  261

\bibitem[{{Asplund} {et~al.}(2009){Asplund}, {Grevesse}, {Sauval}, \&
  {Scott}}]{Asplund2009}
{Asplund}, M., {Grevesse}, N., {Sauval}, A.~J., \& {Scott}, P. 2009, \araa, 47,
  481

\bibitem[{{Bard} {et~al.}(1991){Bard}, {Kock}, \& {Kock}}]{BKK}
{Bard}, A., {Kock}, A., \& {Kock}, M. 1991, Astron. and Astrophys., 248, 315,
  (BKK)

\bibitem[{{Bard} \& {Kock}(1994)}]{BK}
{Bard}, A. \& {Kock}, M. 1994, Astron. and Astrophys., 282, 1014, (BK)

\bibitem[{{Bastian} {et~al.}(2015){Bastian}, {Cabrera-Ziri}, \&
  {Salaris}}]{Bastian2015}
{Bastian}, N., {Cabrera-Ziri}, I., \& {Salaris}, M. 2015, \mnras, 449, 3333

\bibitem[{{Bastian} {et~al.}(2013){Bastian}, {Lamers}, {de Mink}, {Longmore},
  {Goodwin}, \& {Gieles}}]{Bastian2013}
{Bastian}, N., {Lamers}, H.~J.~G.~L.~M., {de Mink}, S.~E., {et~al.} 2013,
  \mnras, 436, 2398

\bibitem[{{Bedin} {et~al.}(2000){Bedin}, {Piotto}, {Zoccali}, {Stetson},
  {Saviane}, {Cassisi}, \& {Bono}}]{Bedin2000}
{Bedin}, L.~R., {Piotto}, G., {Zoccali}, M., {et~al.} 2000, \aap, 363, 159

\bibitem[{{Bergemann} {et~al.}(2012){Bergemann}, {Lind}, {Collet}, {Magic}, \&
  {Asplund}}]{Bergemann2012}
{Bergemann}, M., {Lind}, K., {Collet}, R., {Magic}, Z., \& {Asplund}, M. 2012,
  \mnras, 427, 27

\bibitem[{{Blackwell} {et~al.}(1980){Blackwell}, {Shallis}, \&
  {Simmons}}]{BSScor}
{Blackwell}, D.~E., {Shallis}, M.~J., \& {Simmons}, G.~J. 1980, Astron. and
  Astrophys., 81, 340, (BSScor)

\bibitem[{{Campbell} {et~al.}(2013){Campbell}, {D'Orazi}, {Yong},
  {Constantino}, {Lattanzio}, {Stancliffe}, {Angelou}, {Wylie-de Boer}, \&
  {Grundahl}}]{Campbell2013}
{Campbell}, S.~W., {D'Orazi}, V., {Yong}, D., {et~al.} 2013, \nat, 498, 198

\bibitem[{{Cardelli} {et~al.}(1989){Cardelli}, {Clayton}, \&
  {Mathis}}]{Cardelli1989}
{Cardelli}, J.~A., {Clayton}, G.~C., \& {Mathis}, J.~S. 1989, \apj, 345, 245

\bibitem[{{Carretta}(2014)}]{Carretta2014b}
{Carretta}, E. 2014, \apjl, 795, L28

\bibitem[{{Carretta}(2015)}]{Carretta2015ApJ}
{Carretta}, E. 2015, \apj, 810, 148

\bibitem[{{Carretta} {et~al.}(2004){Carretta}, {Bragaglia}, \&
  {Cacciari}}]{Carretta2004}
{Carretta}, E., {Bragaglia}, A., \& {Cacciari}, C. 2004, \apjl, 610, L25

\bibitem[{{Carretta} {et~al.}(2009{\natexlab{a}}){Carretta}, {Bragaglia},
  {Gratton}, \& {Lucatello}}]{Carretta2009b}
{Carretta}, E., {Bragaglia}, A., {Gratton}, R., \& {Lucatello}, S.
  2009{\natexlab{a}}, \aap, 505, 139

\bibitem[{{Carretta} {et~al.}(2014){Carretta}, {Bragaglia}, {Gratton},
  {D'Orazi}, {Lucatello}, {Momany}, {Sollima}, {Bellazzini}, {Catanzaro}, \&
  {Leone}}]{Carretta2014a}
{Carretta}, E., {Bragaglia}, A., {Gratton}, R.~G., {et~al.} 2014, \aap, 564,
  A60

\bibitem[{{Carretta} {et~al.}(2015){Carretta}, {Bragaglia}, {Gratton},
  {D'Orazi}, {Lucatello}, {Sollima}, {Momany}, {Catanzaro}, \&
  {Leone}}]{Carretta2015}
{Carretta}, E., {Bragaglia}, A., {Gratton}, R.~G., {et~al.} 2015, \aap, 578,
  A116

\bibitem[{{Carretta} {et~al.}(2006){Carretta}, {Bragaglia}, {Gratton}, {Leone},
  {Recio-Blanco}, \& {Lucatello}}]{Carretta2006}
{Carretta}, E., {Bragaglia}, A., {Gratton}, R.~G., {et~al.} 2006, \aap, 450,
  523

\bibitem[{{Carretta} {et~al.}(2009{\natexlab{b}}){Carretta}, {Bragaglia},
  {Gratton}, {Lucatello}, {Catanzaro}, {Leone}, {Bellazzini}, {Claudi},
  {D'Orazi}, {Momany}, {Ortolani}, {Pancino}, {Piotto}, {Recio-Blanco}, \&
  {Sabbi}}]{Carretta2009a}
{Carretta}, E., {Bragaglia}, A., {Gratton}, R.~G., {et~al.} 2009{\natexlab{b}},
  \aap, 505, 117

\bibitem[{{Carretta} {et~al.}(2010){Carretta}, {Bragaglia}, {Gratton},
  {Recio-Blanco}, {Lucatello}, {D'Orazi}, \& {Cassisi}}]{Carretta2010}
{Carretta}, E., {Bragaglia}, A., {Gratton}, R.~G., {et~al.} 2010, \aap, 516,
  A55

\bibitem[{{Carretta} {et~al.}(2005){Carretta}, {Gratton}, {Lucatello},
  {Bragaglia}, \& {Bonifacio}}]{Carretta2005}
{Carretta}, E., {Gratton}, R.~G., {Lucatello}, S., {Bragaglia}, A., \&
  {Bonifacio}, P. 2005, \aap, 433, 597

\bibitem[{{Cassisi} \& {Salaris}(2014)}]{Cassisi2014a}
{Cassisi}, S. \& {Salaris}, M. 2014, \aap, 563, A10

\bibitem[{{Cassisi} {et~al.}(2014){Cassisi}, {Salaris}, {Pietrinferni}, {Vink},
  \& {Monelli}}]{Cassisi2014b}
{Cassisi}, S., {Salaris}, M., {Pietrinferni}, A., {Vink}, J.~S., \& {Monelli},
  M. 2014, \aap, 571, A81

\bibitem[{{Chantereau} {et~al.}(2015){Chantereau}, {Charbonnel}, \&
  {Decressin}}]{Chantereau2015}
{Chantereau}, W., {Charbonnel}, C., \& {Decressin}, T. 2015, \aap, 578, A117

 \bibitem[{{Chantereau} {et~al.}(2016){Chantereau}, {Charbonnel}, \&
   {Meynet}}]{Chantereau2016}
 {Chantereau}, W., {Charbonnel}, C., \& {Meynet}, G. 2016, \aap, in preparation
 
\bibitem[{{Charbonnel} {et~al.}(2013){Charbonnel}, {Chantereau}, {Decressin},
  {Meynet}, \& {Schaerer}}]{Charbonnel2013}
{Charbonnel}, C., {Chantereau}, W., {Decressin}, T., {Meynet}, G., \&
  {Schaerer}, D. 2013, \aap, 557, L17
  
\bibitem[{{Charbonnel} \& {Chantereau}(2016)}]{CCWC2016}
{Charbonnel}, C. \& {Chantereau}, W. 2016, \aap, 586, A21

\bibitem[{{Cordero} {et~al.}(2014){Cordero}, {Pilachowski}, {Johnson},
  {McDonald}, {Zijlstra}, \& {Simmerer}}]{Cordero2014}
{Cordero}, M.~J., {Pilachowski}, C.~A., {Johnson}, C.~I., {et~al.} 2014, \apj,
  780, 94

\bibitem[{{de Mink} {et~al.}(2009){de Mink}, {Pols}, {Langer}, \&
  {Izzard}}]{deMink2009}
{de Mink}, S.~E., {Pols}, O.~R., {Langer}, N., \& {Izzard}, R.~G. 2009, \aap,
  507, L1

\bibitem[{{Decressin} {et~al.}(2007{\natexlab{a}}){Decressin}, {Charbonnel}, \&
  {Meynet}}]{Decressin2007b}
{Decressin}, T., {Charbonnel}, C., \& {Meynet}, G. 2007{\natexlab{a}}, \aap,
  475, 859

\bibitem[{{Decressin} {et~al.}(2007{\natexlab{b}}){Decressin}, {Meynet},
  {Charbonnel}, {Prantzos}, \& {Ekstr{\"o}m}}]{Decressin2007a}
{Decressin}, T., {Meynet}, G., {Charbonnel}, C., {Prantzos}, N., \&
  {Ekstr{\"o}m}, S. 2007{\natexlab{b}}, \aap, 464, 1029

\bibitem[{{Denissenkov} \& {Hartwick}(2014)}]{Denissenkov2014}
{Denissenkov}, P.~A. \& {Hartwick}, F.~D.~A. 2014, \mnras, 437, L21

\bibitem[{{D'Ercole} {et~al.}(2010){D'Ercole}, {D'Antona}, {Ventura},
  {Vesperini}, \& {McMillan}}]{DErcole2010}
{D'Ercole}, A., {D'Antona}, F., {Ventura}, P., {Vesperini}, E., \& {McMillan},
  S.~L.~W. 2010, \mnras, 407, 854

\bibitem[{{Dobrovolskas} {et~al.}(2014){Dobrovolskas}, {Ku{\v c}inskas},
  {Bonifacio}, {Korotin}, {Steffen}, {Sbordone}, {Caffau}, {Ludwig}, {Royer},
  \& {Prakapavi{\v c}ius}}]{Dobrovolskas2014}
{Dobrovolskas}, V., {Ku{\v c}inskas}, A., {Bonifacio}, P., {et~al.} 2014, \aap,
  565, A121

\bibitem[{{Doherty} {et~al.}(2014){Doherty}, {Gil-Pons}, {Lau}, {Lattanzio}, \&
  {Siess}}]{Doherty2014}
{Doherty}, C.~L., {Gil-Pons}, P., {Lau}, H.~H.~B., {Lattanzio}, J.~C., \&
  {Siess}, L. 2014, \mnras, 437, 195

\bibitem[{{D'Orazi} {et~al.}(2010){D'Orazi}, {Lucatello}, {Gratton},
  {Bragaglia}, {Carretta}, {Shen}, \& {Zaggia}}]{DOrazi2010}
{D'Orazi}, V., {Lucatello}, S., {Gratton}, R., {et~al.} 2010, \apjl, 713, L1

\bibitem[{{Forestini} \& {Charbonnel}(1997)}]{Forestini1997}
{Forestini}, M. \& {Charbonnel}, C. 1997, \aaps, 123

\bibitem[{{Fuhr} {et~al.}(1988){Fuhr}, {Martin}, \& {Wiese}}]{FMW}
{Fuhr}, J.~R., {Martin}, G.~A., \& {Wiese}, W.~L. 1988, Journal of Physical and
  Chemical Reference Data, Volume 17, Suppl.~4.~New York: American Institute of
  Physics (AIP) and American Chemical Society, 1988, 17, (FMW)

\bibitem[{{Garc{\'{\i}}a-Hern{\'a}ndez}
  {et~al.}(2015){Garc{\'{\i}}a-Hern{\'a}ndez}, {M{\'e}sz{\'a}ros}, {Monelli},
  {Cassisi}, {Stetson}, {Zamora}, {Shetrone}, \&
  {Lucatello}}]{GarciaHernandez2015}
{Garc{\'{\i}}a-Hern{\'a}ndez}, D.~A., {M{\'e}sz{\'a}ros}, S., {Monelli}, M.,
  {et~al.} 2015, \apjl, 815, L4

\bibitem[{{Gehren} {et~al.}(2004){Gehren}, {Liang}, {Shi}, {Zhang}, \&
  {Zhao}}]{Gehren2004}
{Gehren}, T., {Liang}, Y.~C., {Shi}, J.~R., {Zhang}, H.~W., \& {Zhao}, G. 2004,
  \aap, 413, 1045

\bibitem[{{Gratton} {et~al.}(2001){Gratton}, {Bonifacio}, {Bragaglia},
  {Carretta}, {Castellani}, {Centurion}, {Chieffi}, {Claudi}, {Clementini},
  {D'Antona}, {Desidera}, {Fran{\c c}ois}, {Grundahl}, {Lucatello}, {Molaro},
  {Pasquini}, {Sneden}, {Spite}, \& {Straniero}}]{Gratton2001}
{Gratton}, R.~G., {Bonifacio}, P., {Bragaglia}, A., {et~al.} 2001, \aap, 369,
  87

\bibitem[{{Gratton} {et~al.}(2003){Gratton}, {Bragaglia}, {Carretta},
  {Clementini}, {Desidera}, {Grundahl}, \& {Lucatello}}]{Gratton2003}
{Gratton}, R.~G., {Bragaglia}, A., {Carretta}, E., {et~al.} 2003, \aap, 408,
  529

\bibitem[{{Gratton} {et~al.}(2012{\natexlab{a}}){Gratton}, {Carretta}, \&
  {Bragaglia}}]{Gratton2012AARv}
{Gratton}, R.~G., {Carretta}, E., \& {Bragaglia}, A. 2012{\natexlab{a}}, \aapr,
  20, 50

\bibitem[{{Gratton} {et~al.}(2012{\natexlab{b}}){Gratton}, {Lucatello},
  {Carretta}, {Bragaglia}, {D'Orazi}, {Al Momany}, {Sollima}, {Salaris}, \&
  {Cassisi}}]{Gratton2012}
{Gratton}, R.~G., {Lucatello}, S., {Carretta}, E., {et~al.} 2012{\natexlab{b}},
  \aap, 539, A19

\bibitem[{{Gratton} {et~al.}(2011){Gratton}, {Lucatello}, {Carretta},
  {Bragaglia}, {D'Orazi}, \& {Momany}}]{Gratton2011}
{Gratton}, R.~G., {Lucatello}, S., {Carretta}, E., {et~al.} 2011, \aap, 534,
  A123

\bibitem[{{Gratton} {et~al.}(2013){Gratton}, {Lucatello}, {Sollima},
  {Carretta}, {Bragaglia}, {Momany}, {D'Orazi}, {Cassisi}, {Pietrinferni}, \&
  {Salaris}}]{Gratton2013}
{Gratton}, R.~G., {Lucatello}, S., {Sollima}, A., {et~al.} 2013, \aap, 549, A41

\bibitem[{{Gratton} {et~al.}(2014){Gratton}, {Lucatello}, {Sollima},
  {Carretta}, {Bragaglia}, {Momany}, {D'Orazi}, {Cassisi}, \&
  {Salaris}}]{Gratton2014}
{Gratton}, R.~G., {Lucatello}, S., {Sollima}, A., {et~al.} 2014, \aap, 563, A13

\bibitem[{{Gratton} {et~al.}(2015){Gratton}, {Lucatello}, {Sollima},
  {Carretta}, {Bragaglia}, {Momany}, {D'Orazi}, {Salaris}, {Cassisi}, \&
  {Stetson}}]{Gratton2015}
{Gratton}, R.~G., {Lucatello}, S., {Sollima}, A., {et~al.} 2015, \aap, 573, A92

\bibitem[{{Gustafsson} {et~al.}(2008){Gustafsson}, {Edvardsson}, {Eriksson},
  {J{\o}rgensen}, {Nordlund}, \& {Plez}}]{Gustafsson2008}
{Gustafsson}, B., {Edvardsson}, B., {Eriksson}, K., {et~al.} 2008, \aap, 486,
  951

\bibitem[{{Harris}(1996)}]{Harris1996}
{Harris}, W.~E. 1996, \aj, 112, 1487

\bibitem[{{Izzard} {et~al.}(2013){Izzard}, {de Mink}, {Pols}, {Langer}, {Sana},
  \& {de Koter}}]{Izzard2013}
{Izzard}, R.~G., {de Mink}, S.~E., {Pols}, O.~R., {et~al.} 2013, \memsai, 84,
  171

\bibitem[{{Johnson} {et~al.}(2015){Johnson}, {McDonald}, {Pilachowski},
  {Mateo}, {Bailey}, {Cordero}, {Zijlstra}, {Crane}, {Olszewski}, {Shectman},
  \& {Thompson}}]{Johnson2015}
{Johnson}, C.~I., {McDonald}, I., {Pilachowski}, C.~A., {et~al.} 2015, \aj,
  149, 71

\bibitem[{{Johnson} \& {Pilachowski}(2012)}]{Johnson2012}
{Johnson}, C.~I. \& {Pilachowski}, C.~A. 2012, \apjl, 754, L38

\bibitem[{{Kausch} {et~al.}(2015){Kausch}, {Noll}, {Smette}, {Kimeswenger},
  {Barden}, {Szyszka}, {Jones}, {Sana}, {Horst}, \& {Kerber}}]{Kausch2015}
{Kausch}, W., {Noll}, S., {Smette}, A., {et~al.} 2015, \aap, 576, A78

\bibitem[{{Kraft} {et~al.}(1992){Kraft}, {Sneden}, {Langer}, \&
  {Prosser}}]{Kraft1992}
{Kraft}, R.~P., {Sneden}, C., {Langer}, G.~E., \& {Prosser}, C.~F. 1992, \aj,
  104, 645

\bibitem[{{Kraft} {et~al.}(1993){Kraft}, {Sneden}, {Langer}, \&
  {Shetrone}}]{Kraft1993}
{Kraft}, R.~P., {Sneden}, C., {Langer}, G.~E., \& {Shetrone}, M.~D. 1993, \aj,
  106, 1490

\bibitem[{{Kraft} {et~al.}(1995){Kraft}, {Sneden}, {Langer}, {Shetrone}, \&
  {Bolte}}]{Kraft1995}
{Kraft}, R.~P., {Sneden}, C., {Langer}, G.~E., {Shetrone}, M.~D., \& {Bolte},
  M. 1995, \aj, 109, 2586

\bibitem[{{Krause} {et~al.}(2013){Krause}, {Charbonnel}, {Decressin}, {Meynet},
  \& {Prantzos}}]{Krause2013}
{Krause}, M., {Charbonnel}, C., {Decressin}, T., {Meynet}, G., \& {Prantzos},
  N. 2013, \aap, 552, A121

\bibitem[{{Krause} {et~al.}(2016){Krause}, {Charbonnel}, {Bastian}, \&
  {Diehl}}]{Krause2016}
{Krause}, M.~G.~H., {Charbonnel}, C., {Bastian}, N., \& {Diehl}, R. 2016, \aap,
  587, A53

\bibitem[{{Kupka} {et~al.}(2000){Kupka}, {Ryabchikova}, {Piskunov}, {Stempels},
  \& {Weiss}}]{Kupka2000VALD}
{Kupka}, F.~G., {Ryabchikova}, T.~A., {Piskunov}, N.~E., {Stempels}, H.~C., \&
  {Weiss}, W.~W. 2000, Baltic Astronomy, 9, 590

\bibitem[{{Kurucz}(2007)}]{K07}
{Kurucz}, R.~L. 2007, Robert L. Kurucz on-line database of observed and
  predicted atomic transitions

\bibitem[{{Kurucz}(2013)}]{K13}
{Kurucz}, R.~L. 2013, Robert L. Kurucz on-line database of observed and
  predicted atomic transitions

\bibitem[{{Kurucz} \& {Peytremann}(1975)}]{KP}
{Kurucz}, R.~L. \& {Peytremann}, E. 1975, SAO Special Report, 362, 1, (KP)

\bibitem[{{Lapenna} {et~al.}(2015){Lapenna}, {Mucciarelli}, {Ferraro},
  {Origlia}, {Lanzoni}, {Massari}, \& {Dalessandro}}]{Lapenna2015}
{Lapenna}, E., {Mucciarelli}, A., {Ferraro}, F.~R., {et~al.} 2015, \apj, 813,
  97

\bibitem[{{Lind} {et~al.}(2011){Lind}, {Asplund}, {Barklem}, \&
  {Belyaev}}]{Lind2011}
{Lind}, K., {Asplund}, M., {Barklem}, P.~S., \& {Belyaev}, A.~K. 2011, \aap,
  528, A103

\bibitem[{{Lind} {et~al.}(2012){Lind}, {Bergemann}, \& {Asplund}}]{Lind2012}
{Lind}, K., {Bergemann}, M., \& {Asplund}, M. 2012, \mnras, 427, 50

\bibitem[{{Lind} {et~al.}(2009){Lind}, {Primas}, {Charbonnel}, {Grundahl}, \&
  {Asplund}}]{Lind2009}
{Lind}, K., {Primas}, F., {Charbonnel}, C., {Grundahl}, F., \& {Asplund}, M.
  2009, \aap, 503, 545

\bibitem[{{Maeder} \& {Meynet}(2006)}]{MaederMeynet2006}
{Maeder}, A. \& {Meynet}, G. 2006, \aap, 448, L37

\bibitem[{{Massari} {et~al.}(2016){Massari}, {Fiorentino}, {McConnachie},
  {Bono}, {Dall'Ora}, {Ferraro}, {Iannicola}, {Stetson}, {Turri}, \&
  {Tolstoy}}]{Massari2016}
{Massari}, D., {Fiorentino}, G., {McConnachie}, A., {et~al.} 2016, \aap, 586,
  A51

\bibitem[{{Milone} {et~al.}(2015{\natexlab{a}}){Milone}, {Marino}, {Piotto},
  {Bedin}, {Anderson}, {Renzini}, {King}, {Bellini}, {Brown}, {Cassisi},
  {D'Antona}, {Jerjen}, {Nardiello}, {Salaris}, {Marel}, {Vesperini}, {Yong},
  {Aparicio}, {Sarajedini}, \& {Zoccali}}]{Milone2015a}
{Milone}, A.~P., {Marino}, A.~F., {Piotto}, G., {et~al.} 2015{\natexlab{a}},
  \mnras, 447, 927

\bibitem[{{Milone} {et~al.}(2015{\natexlab{b}}){Milone}, {Marino}, {Piotto},
  {Renzini}, {Bedin}, {Anderson}, {Cassisi}, {D'Antona}, {Bellini}, {Jerjen},
  {Pietrinferni}, \& {Ventura}}]{Milone2015b}
{Milone}, A.~P., {Marino}, A.~F., {Piotto}, G., {et~al.} 2015{\natexlab{b}},
  \apj, 808, 51

\bibitem[{{Monaco} {et~al.}(2012){Monaco}, {Villanova}, {Bonifacio}, {Caffau},
  {Geisler}, {Marconi}, {Momany}, \& {Ludwig}}]{Monaco2012}
{Monaco}, L., {Villanova}, S., {Bonifacio}, P., {et~al.} 2012, \aap, 539, A157

\bibitem[{{Nardiello} {et~al.}(2015){Nardiello}, {Piotto}, {Milone}, {Marino},
  {Bedin}, {Anderson}, {Aparicio}, {Bellini}, {Cassisi}, {D'Antona}, {Hidalgo},
  {Ortolani}, {Pietrinferni}, {Renzini}, {Salaris}, {Marel}, \&
  {Vesperini}}]{Nardiello2015}
{Nardiello}, D., {Piotto}, G., {Milone}, A.~P., {et~al.} 2015, \mnras, 451, 312

\bibitem[{{O'Brian} {et~al.}(1991){O'Brian}, {Wickliffe}, {Lawler}, {Whaling},
  \& {Brault}}]{BWL}
{O'Brian}, T.~R., {Wickliffe}, M.~E., {Lawler}, J.~E., {Whaling}, W., \&
  {Brault}, J.~W. 1991, Journal of the Optical Society of America B Optical
  Physics, 8, 1185, (BWL)

\bibitem[{{Pancino} {et~al.}(2011){Pancino}, {Mucciarelli}, {Sbordone},
  {Bellazzini}, {Pasquini}, {Monaco}, \& {Ferraro}}]{Pancino2011}
{Pancino}, E., {Mucciarelli}, A., {Sbordone}, L., {et~al.} 2011, \aap, 527, A18

\bibitem[{{Pancino} {et~al.}(2010){Pancino}, {Rejkuba}, {Zoccali}, \&
  {Carrera}}]{Pancino2010}
{Pancino}, E., {Rejkuba}, M., {Zoccali}, M., \& {Carrera}, R. 2010, \aap, 524,
  A44

\bibitem[{{Pasquini} {et~al.}(2003){Pasquini}, {Alonso}, {Avila}, {Barriga},
  {Biereichel}, {Buzzoni}, {Cavadore}, {Cumani}, {Dekker}, {Delabre}, {Kaufer},
  {Kotzlowski}, {Hill}, {Lizon}, {Nees}, {Santin}, {Schmutzer}, {Kesteren}, \&
  {Zoccali}}]{Pasquini2003}
{Pasquini}, L., {Alonso}, J., {Avila}, G., {et~al.} 2003, in Society of
  Photo-Optical Instrumentation Engineers (SPIE) Conference Series, Vol. 4841,
  Instrument Design and Performance for Optical/Infrared Ground-based
  Telescopes, ed. M.~{Iye} \& A.~F.~M. {Moorwood}, 1682--1693

\bibitem[{{Pilachowski} {et~al.}(1996){Pilachowski}, {Sneden}, {Kraft}, \&
  {Langer}}]{Pilachowski1996}
{Pilachowski}, C.~A., {Sneden}, C., {Kraft}, R.~P., \& {Langer}, G.~E. 1996,
  \aj, 112, 545

\bibitem[{{Piotto} {et~al.}(2007){Piotto}, {Bedin}, {Anderson}, {King},
  {Cassisi}, {Milone}, {Villanova}, {Pietrinferni}, \& {Renzini}}]{Piotto2007}
{Piotto}, G., {Bedin}, L.~R., {Anderson}, J., {et~al.} 2007, \apjl, 661, L53

\bibitem[{{Piotto} {et~al.}(2012){Piotto}, {Milone}, {Anderson}, {Bedin},
  {Bellini}, {Cassisi}, {Marino}, {Aparicio}, \& {Nascimbeni}}]{Piotto2012}
{Piotto}, G., {Milone}, A.~P., {Anderson}, J., {et~al.} 2012, \apj, 760, 39

\bibitem[{{Piotto} {et~al.}(2015){Piotto}, {Milone}, {Bedin}, {Anderson},
  {King}, {Marino}, {Nardiello}, {Aparicio}, {Barbuy}, {Bellini}, {Brown},
  {Cassisi}, {Cool}, {Cunial}, {Dalessandro}, {D'Antona}, {Ferraro}, {Hidalgo},
  {Lanzoni}, {Monelli}, {Ortolani}, {Renzini}, {Salaris}, {Sarajedini}, {van
  der Marel}, {Vesperini}, \& {Zoccali}}]{Piotto2015}
{Piotto}, G., {Milone}, A.~P., {Bedin}, L.~R., {et~al.} 2015, \aj, 149, 91

\bibitem[{{Piskunov} {et~al.}(1995){Piskunov}, {Kupka}, {Ryabchikova}, {Weiss},
  \& {Jeffery}}]{Piskunov1995VLAD}
{Piskunov}, N.~E., {Kupka}, F., {Ryabchikova}, T.~A., {Weiss}, W.~W., \&
  {Jeffery}, C.~S. 1995, \aaps, 112, 525

\bibitem[{{Prantzos} \& {Charbonnel}(2006)}]{PrantzosCharbonnel2006}
{Prantzos}, N. \& {Charbonnel}, C. 2006, \aap, 458, 135

\bibitem[{{Prantzos} {et~al.}(2007){Prantzos}, {Charbonnel}, \&
  {Iliadis}}]{Prantzos2007}
{Prantzos}, N., {Charbonnel}, C., \& {Iliadis}, C. 2007, \aap, 470, 179

\bibitem[{{Ralchenko} {et~al.}(2010){Ralchenko}, {Kramida}, {Reader}, \& {NIST
  ASD Team}}]{NIST10}
{Ralchenko}, Y., {Kramida}, A., {Reader}, J., \& {NIST ASD Team}. 2010, NIST
  Atomic Spectra Database (ver. 4.0.0), [Online].

\bibitem[{{Ram{\'{\i}}rez} \& {Mel{\'e}ndez}(2005)}]{RamirezMelendez2005}
{Ram{\'{\i}}rez}, I. \& {Mel{\'e}ndez}, J. 2005, \apj, 626, 465

\bibitem[{{Renzini} {et~al.}(2015){Renzini}, {D'Antona}, {Cassisi}, {King},
  {Milone}, {Ventura}, {Anderson}, {Bedin}, {Bellini}, {Brown}, {Piotto}, {van
  der Marel}, {Barbuy}, {Dalessandro}, {Hidalgo}, {Marino}, {Ortolani},
  {Salaris}, \& {Sarajedini}}]{Renzini2015}
{Renzini}, A., {D'Antona}, F., {Cassisi}, S., {et~al.} 2015, \mnras, 454, 4197

\bibitem[{{Ryabchikova} {et~al.}(2015){Ryabchikova}, {Piskunov}, {Kurucz},
  {Stempels}, {Heiter}, {Pakhomov}, \& {Barklem}}]{Ryabchikova2015VALD}
{Ryabchikova}, T., {Piskunov}, N., {Kurucz}, R.~L., {et~al.} 2015, \physscr,
  90, 054005

\bibitem[{{Sills} \& {Glebbeek}(2010)}]{Sills2010}
{Sills}, A. \& {Glebbeek}, E. 2010, \mnras, 407, 277

\bibitem[{{Skrutskie} {et~al.}(2006){Skrutskie}, {Cutri}, {Stiening},
  {Weinberg}, {Schneider}, {Carpenter}, {Beichman}, {Capps}, {Chester},
  {Elias}, {Huchra}, {Liebert}, {Lonsdale}, {Monet}, {Price}, {Seitzer},
  {Jarrett}, {Kirkpatrick}, {Gizis}, {Howard}, {Evans}, {Fowler}, {Fullmer},
  {Hurt}, {Light}, {Kopan}, {Marsh}, {McCallon}, {Tam}, {Van Dyk}, \&
  {Wheelock}}]{Skrutskie2006}
{Skrutskie}, M.~F., {Cutri}, R.~M., {Stiening}, R., {et~al.} 2006, \aj, 131,
  1163

\bibitem[{{Smette} {et~al.}(2015){Smette}, {Sana}, {Noll}, {Horst}, {Kausch},
  {Kimeswenger}, {Barden}, {Szyszka}, {Jones}, {Gallenne}, {Vinther},
  {Ballester}, \& {Taylor}}]{Smette2015}
{Smette}, A., {Sana}, H., {Noll}, S., {et~al.} 2015, \aap, 576, A77

\bibitem[{{Sneden} {et~al.}(1994){Sneden}, {Kraft}, {Langer}, {Prosser}, \&
  {Shetrone}}]{Sneden1994}
{Sneden}, C., {Kraft}, R.~P., {Langer}, G.~E., {Prosser}, C.~F., \& {Shetrone},
  M.~D. 1994, \aj, 107, 1773

\bibitem[{{Sneden} {et~al.}(1992){Sneden}, {Kraft}, {Prosser}, \&
  {Langer}}]{Sneden1992}
{Sneden}, C., {Kraft}, R.~P., {Prosser}, C.~F., \& {Langer}, G.~E. 1992, \aj,
  104, 2121

\bibitem[{{Sneden}(1973)}]{Sneden1973}
{Sneden}, C.~A. 1973, PhD thesis, THE UNIVERSITY OF TEXAS AT AUSTIN.

\bibitem[{{Stetson}(2000)}]{Stetson2000}
{Stetson}, P.~B. 2000, \pasp, 112, 925

\bibitem[{{Stetson}(2005)}]{Stetson2005}
{Stetson}, P.~B. 2005, \pasp, 117, 563

\bibitem[{{Stetson} \& {Pancino}(2008)}]{Stetson2008}
{Stetson}, P.~B. \& {Pancino}, E. 2008, \pasp, 120, 1332

\bibitem[{{VandenBerg} {et~al.}(2013){VandenBerg}, {Brogaard}, {Leaman}, \&
  {Casagrande}}]{VandenBerg2013}
{VandenBerg}, D.~A., {Brogaard}, K., {Leaman}, R., \& {Casagrande}, L. 2013,
  \apj, 775, 134

\bibitem[{{Ventura} \& {D'Antona}(2011)}]{Ventura2011}
{Ventura}, P. \& {D'Antona}, F. 2011, \mnras, 410, 2760

\bibitem[{{Ventura} {et~al.}(2001){Ventura}, {D'Antona}, {Mazzitelli}, \&
  {Gratton}}]{Ventura2001}
{Ventura}, P., {D'Antona}, F., {Mazzitelli}, I., \& {Gratton}, R. 2001, \apjl,
  550, L65

\bibitem[{{Ventura} {et~al.}(2013){Ventura}, {Di Criscienzo}, {Carini}, \&
  {D'Antona}}]{Ventura2013}
{Ventura}, P., {Di Criscienzo}, M., {Carini}, R., \& {D'Antona}, F. 2013,
  \mnras, 431, 3642

\bibitem[{{Villanova} \& {Geisler}(2011)}]{Villanova2011}
{Villanova}, S. \& {Geisler}, D. 2011, \aap, 535, A31

\bibitem[{{Villanova} {et~al.}(2009){Villanova}, {Piotto}, \&
  {Gratton}}]{Villanova2009}
{Villanova}, S., {Piotto}, G., \& {Gratton}, R.~G. 2009, \aap, 499, 755

\bibitem[{{Yong} {et~al.}(2013){Yong}, {Mel{\'e}ndez}, {Grundahl}, {Roederer},
  {Norris}, {Milone}, {Marino}, {Coelho}, {McArthur}, {Lind}, {Collet}, \&
  {Asplund}}]{Yong2013}
{Yong}, D., {Mel{\'e}ndez}, J., {Grundahl}, F., {et~al.} 2013, \mnras, 434,
  3542

\end{thebibliography}


\begin{appendix} 
\section{Spectroscopic stellar parameters} 

We selected ten stars (4 AGB and 6 RGB) representative of our GIRAFFE sample (i.e. covering their entire temperature ranges). We then derived the 
(excitation) temperature, by requiring a null slope between the \ion{Fe}{i} abundances and the excitation potential of the individual \ion{Fe}{i} lines 
(the so-called excitation equilibrium) and the surface gravity, by requiring \ion{Fe}{i} and \ion{Fe}{ii} lines to give the same abundance (the so-called 
ionization equilibrium). Stellar metallicity and microturbulent velocities were determined, as described in Sect. \ref{section:Metallicity}. We note that 
by imposing the ionisation balance between \ion{Fe}{i}$_{NLTE}$ and \ion{Fe}{ii} (i.e. taking somehow into account a mean NLTE correction to \ion{Fe}{i} 
of $\simeq$0.04\,dex) only the gravity and the \ion{Fe}{ii} abundance are slightly affected ($\log g$ by $+0.08$\,dex $-$ from LTE to NLTE $-$ and 
\ion{Fe}{ii} by $+0.05$\,dex). 

As far as the temperature is concerned, we find a mean difference that is smaller than our estimated error on this parameter 
($T_\mathrm{spec} - T_\mathrm{phot}=-26\pm56\,\mathrm{K}$). The standard deviation around the mean value, however, hints to possibly important dependencies 
on the type of star: the temperature difference is slightly larger in the AGB stars ($-$34\,K) than in the RGB ones ($-$26\,K); cool stars (AGB and RGB) 
show positive differences, while the hotter ones tend to be negative. This is observed also for all the other parameters (surface gravity and microturbulent 
velocity) and abundances (\ion{Fe}{i}, \ion{Fe}{ii}, \ion{Na}{i})\footnote{$\xi_\mathrm{t,spec} - \xi_\mathrm{t,phot}=-0.02\pm0.05\,\mathrm{km\,s}^{-1}$ 
and $\mathrm{[\ion{Fe}{i}/H]}_\mathrm{spec} - \mathrm{[\ion{Fe}{i}/H]}_\mathrm{phot}=-0.03\pm0.06\,\mathrm{dex}$. See the text for other parameters and 
abundances.}, with the surface gravity being the most affected quantity.  

On average, the ionisation equilibrium requirement (omitting the NLTE effect on \ion{Fe}{i}) decreases $\log g$ by $-0.14\pm0.24\,\mathrm{dex}$ 
and the \ion{Fe}{ii} abundance by $-0.05\pm0.09\,\mathrm{dex}$ with respect to the photometry-based result. The differences on $\log g$ and \ion{Fe}{ii} 
decrease to $-0.06\pm0.2\,\mathrm{dex}$ and $-0.01\pm0.09\,\mathrm{dex}$ respectively, when we apply NLTE corrections to the \ion{Fe}{i} values before 
seeking for the ionisation balance. The \ion{Fe}{i} abundances differ by $-0.03\pm0.6\,\mathrm{dex}$ between the spectroscopic (lower) and photometric 
results, independently of the NLTE correction. 

When deriving the \element{Na} abundances (we used the EWs of the 6154$-$6160\,\AA\ \element{Na} doublet for the purpose of this test) with the 
spectroscopic set of stellar parameters, we find a negligible difference on the derived [\element{Na}/\element{H}] ratios ($\Delta=-0.02$, 
rms$=0.04$\,$\mathrm{dex}$, with the spectroscopic one being lower). This is much smaller than their associated errors and the difference between 
AGB and RGB samples is negligible. 

In summary, the effect of applying a photometric or spectroscopic method to derive stellar parameters and abundances is not significant and 
the resulting sets of abundances agree well within the associated errors. 

\end{appendix}

\end{document}